\documentclass{PoS}

\usepackage{bm}
\usepackage{amsmath,dcolumn}
\usepackage{rotating}
\usepackage{mathtools}

\newcolumntype{d}[1]{D{.}{.}{#1} }

\title{Rare \textit{B} decays using lattice QCD form factors}

\ShortTitle{Rare $B$ decays using lattice QCD form factors}

\author{R.~R.~Horgan\\
  Department of Applied Mathematics and Theoretical Physics, University of
  Cambridge, Cambridge CB3 0WA, UK\\
  E-mail: \email{R.R.Horgan@damtp.cam.ac.uk}}

\author{Z.~Liu\\
  Institute of High Energy Physics and Theoretical Physics Center for Science
  Facilities, Chinese Academy of Sciences, Beijing 100049, China\\
  E-mail: \email{liuzf@ihep.ac.cn}}

\author{S.~Meinel\\
  Department of Physics, University of Arizona, Tucson, AZ, 85721, USA\\
  and RIKEN BNL Research Center, Brookhaven National Laboratory, Upton NY 
  11973, USA \\
        E-mail: \email{smeinel@email.arizona.edu}}

\author{\speaker{M. Wingate}
  \\
  Department of Applied Mathematics and Theoretical Physics, University of
  Cambridge, Cambridge CB3 0WA, UK\\
  E-mail: \email{M.Wingate@damtp.cam.ac.uk}}

\abstract{In this write-up we review and update our recent lattice
  QCD calculation of $B \to K^*$, $B_s \to \phi$, and $B_s \to K^*$ form
  factors \cite{Horgan:2013hoa}.  These unquenched calculations, performed in
  the low-recoil kinematic regime, provide a significant improvement over the
  use of extrapolated light cone sum rule results.  The fits presented here
  include further kinematic constraints and estimates of additional
  correlations between the different form factor shape parameters.  We use
  these form factors along with Standard Model determinations of Wilson
  coefficients to give Standard Model predictions for several observables
  \cite{Horgan:2013pva}.  The modest improvements to the form factor fits lead
  to improved determinations of $F_L$, the fraction of longitudinally
  polarized vector mesons, but have little effect on most other observables.}

\FullConference{The 32nd International Symposium on Lattice Field Theory\\
		 23-28 June, 2014\\
		 Columbia University, New York, NY}

\begin{document}

\section{Introduction}

Measurements of the flavor-changing, neutral current (FCNC) decay
$b\to s$ are rapidly growing in number.  In particular, the
semi-leptonic FCNC decays of $B$ and $B_s$ mesons to the vector mesons,
$K^*$ and $\phi$ respectively, have afforded us the opportunity to
compare several associated observables with theoretical predictions
\cite{Aaij:2013iag,ATLAS:2013ola,Aaij:2013aln,Aaij:2014pli}.  The
experimental data presently show a few deviations from the Standard Model,
including a smaller-than-predicted differential branching fraction in 
low-recoil kinematic bins.  These could be explained by
beyond-the-Standard-Model (BSM) contributions to one or two Wilson
coefficients in the effective $b\to s$ Hamiltonian
\cite{Descotes-Genon:2013wba,Altmannshofer:2013foa,Hambrock:2013zya,Beaujean:2013soa,Horgan:2013pva,Hurth:2014vma,Altmannshofer:2014rta}. However,
none of the analyses claim the Standard Model is an unacceptable
description of the data.  In fact it is not yet certain that all theoretical
uncertainties are under firm control.

The largest theoretical uncertainties in these observables are due to QCD
interactions.  Lattice QCD (LQCD) calculations of the $B_{(s)} \to K^*/\phi$
form factors can reduce a class of these uncertainties, improving upon the
determinations from sum rules.  We recently completed a necessary first step
toward accurate, first-principles calculations of the form factors
\cite{Horgan:2013hoa}.  Below we briefly summarize the results of computing
matrix elements of the full basis of $b\to s$ currents.  The key improvements
made compared to previous lattice studies come through using high statistics,
physical-mass bottom quarks, and $2+1$ flavors of sea quarks.  In
Sec.~\ref{sec:fits} we provide a minor improvement to the method used to fit
the form factor shapes.  We close in Sec.~\ref{sec:open} by touching upon
issues which must be confronted in order to improve the theoretical
determinations of these observables.

\section{Summary of the matrix element calculation}
\label{sec:brief}

The effective Hamiltonian governing $b\to s$ decay is
\begin{equation}
\mathcal{H}_{\mathrm{eff}}^{b\to s} ~=~ -\frac{4 G_F}{\sqrt{2}}
V_{ts}^* V_{tb} \sum_i (C_i O_i + C_i' O_i') \,.
\label{eq:Heff}
\end{equation}
For the following operators in $\mathcal{H}_{\mathrm{eff}}^{b\to s}$,
matrix elements factorize into local hadronic matrix elements,
which are parametrized by seven form factors:
\begin{equation}
O_{\mathrlap{7}^{\vphantom '}}{}^{(}{}'^{)} = \frac{m_b e}{16\pi^2} 
\bar{s}\sigma^{\mu\nu} P_{R(L)} b \, F_{\mu\nu}\,, ~
O_{\mathrlap{9}^{\vphantom '}}{}^{(}{}'^{)} = \frac{e^2}{16\pi^2}\bar{s} 
\gamma^\mu P_{L(R)} b\, \bar\ell \gamma_\mu \ell\,,
~\mbox{and}~
O_{\mathrlap{10}^{\vphantom '}}{}^{(}{}'^{)} = \frac{e^2}{16\pi^2}\bar{s} 
\gamma^\mu P_{L(R)} b\, \bar\ell \gamma_\mu \gamma^5 \ell \,,
\label{eq:ops}
\end{equation}
where $P_{L/R} = \tfrac{1}{2}(1 \mp \gamma^5)$ and $\sigma^{\mu\nu} =
\tfrac{i}{2}[\gamma^\mu,\gamma^\nu]$.  The bulk of this write-up
focuses on the details of the form factor calculation and
consequences; however, we will comment on contributions to $b\to s$
from nonlocal matrix elements and other open issues at the end.

We used a subset of the MILC Collaboration gauge field configurations
which include effects of $2+1$ flavors of $O(a^2)$, tadpole-improved
(asqtad) sea quarks.  We chose three ensembles, allowing for modest tests
of discretization and quark mass effects.  We correctly anticipated
that obtaining a good signal over statistical noise would be a
challenge.  In order to address this issue we used eight light quark
sources on each configuration, obtaining over 30000 estimates of each
correlation function on each of the three ensembles.

The calculations were done with nonrelativistic $b$ quarks, formulated in
the $B_{(s)}$ rest frame and accurate through $O(v^4)$.  The matching of
the effective field theory currents to the physical ones, done to
$O(\alpha_s^2, \alpha_s \Lambda_{\mathrm{QCD}}/m_b,
\Lambda_{\mathrm{QCD}}^2/m_b^2 )$, is the source of the largest controlled
systematic uncertainty.

The momentum of the $K^*$ or $\phi$ meson varied from $0$ to no greater
than $4\pi/L$ ($\approx 1$ GeV) in magnitude, primarily due to signal-to-noise
degradation.  Thus the lattice calculations are done in the low recoil 
regime, i.e.\ at large $q^2$, the lepton-pair invariant mass-squared.

From the imaginary-time correlation functions, we performed
bootstrapped fits to extract numerical estimates of the seven linearly
independent form factors.  The basis we use is $\{V, A_0$, $A_1$,
$A_{12}$, $T_1$, $T_2$, $T_{23}\}$ (see \cite{Horgan:2013hoa} for
definitions).  Once these are determined on each ensemble, for several
values of final state momentum, it remains for us to parametrize the
shape and give results corresponding to the physical limit.

\section{Updated fits}
\label{sec:fits}

This section describes a minor update of our published results
\cite{Horgan:2013hoa}.  We fit each form factor, generically denoted $F(t)$,
using the following parametrization
\begin{equation}
F(t) = \frac{1}{1 - t/(m_{B_{(s)}} + \Delta m^F)^2}\left[ a_0^F\left(1 + 
c_{01}^F \Delta x + c_{01s}^F \Delta x_s\right) + a_1^F z(t;t_0)\right] \,
\label{eq:sse_3par}
\end{equation}
where $\Delta x = (m_\pi^2 - m_{\pi,\mathrm{phys}}^2)/(4\pi f_\pi)^2$
and $\Delta x_s = (m_{\eta_s}^2 - m_{\eta_s,\mathrm{phys}}^2)/(4\pi
f_\pi)^2$.  Thus the parameters $c_{01}^F$ and $c_{01s}^F$ quantify
the dependence of the form factors on the light and strange quark
masses, respectively.  The splitting $\Delta m^F$ between the $B_{(s)}$
mass and the relevant resonance is taken to be a fixed value, but variations
of up to $20\%$ had no effect on the resulting physical form factors.
We determine $c_{01s}^F$ from a combined fit to
$B\to K^*$, $B_s\to\phi$, and $B_s\to K^*$ data \cite{Horgan:2013hoa}
and include it as a Gaussian prior in the subsequent fits to determine
$a_0^F$, $a_1^F$, and $c_{01}^F$.  Physical results are obtained by
using our fit results for $a_0^F$ and $a_1^F$ with $\Delta x = \Delta
x_s = 0$.  We tried several fits which allowed for discretization
effects, but no statistical signal was seen for the corresponding
parameters.

\begin{figure}
\centering
\includegraphics[width=0.329\textwidth]{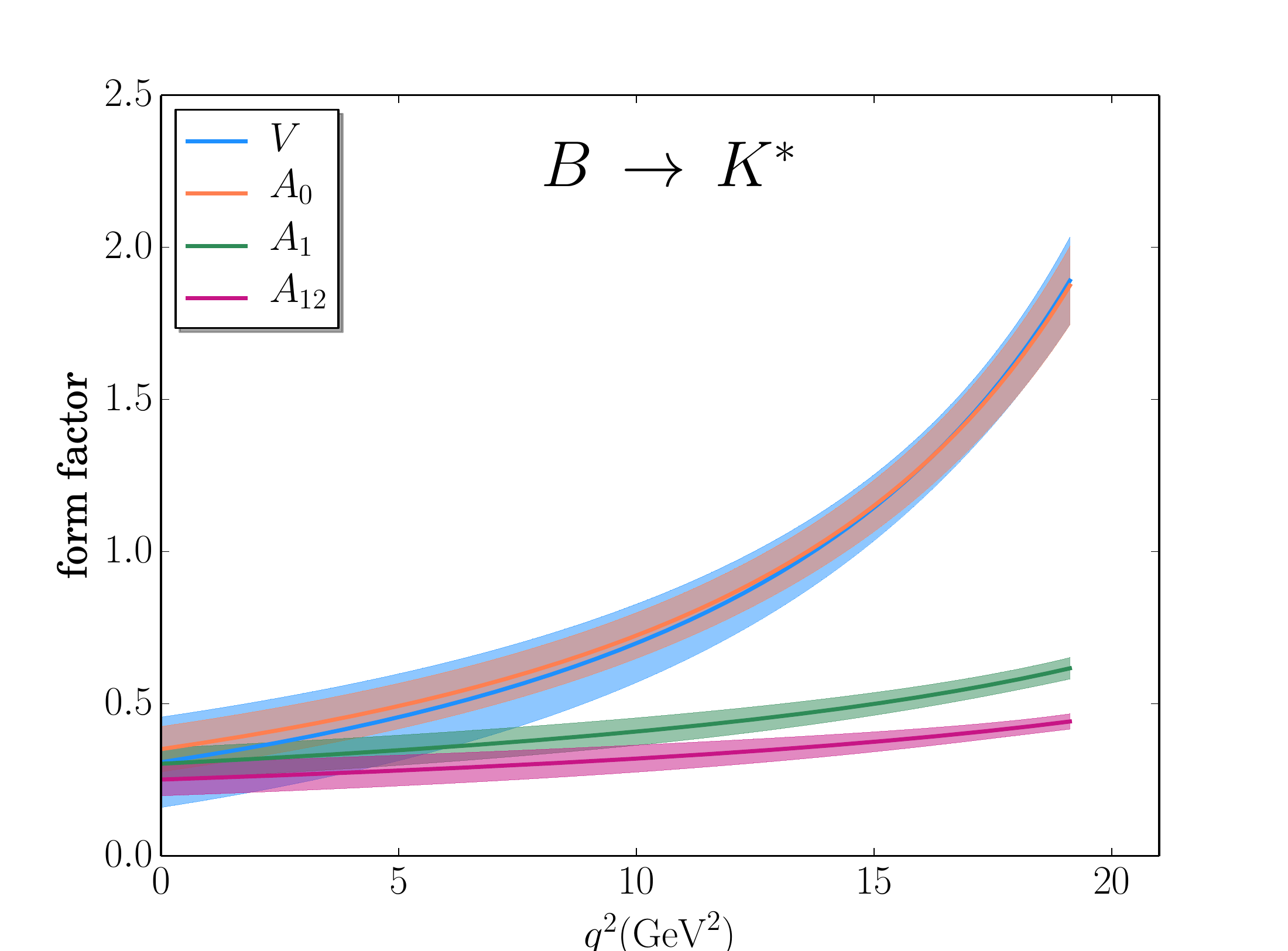}
\includegraphics[width=0.329\textwidth]{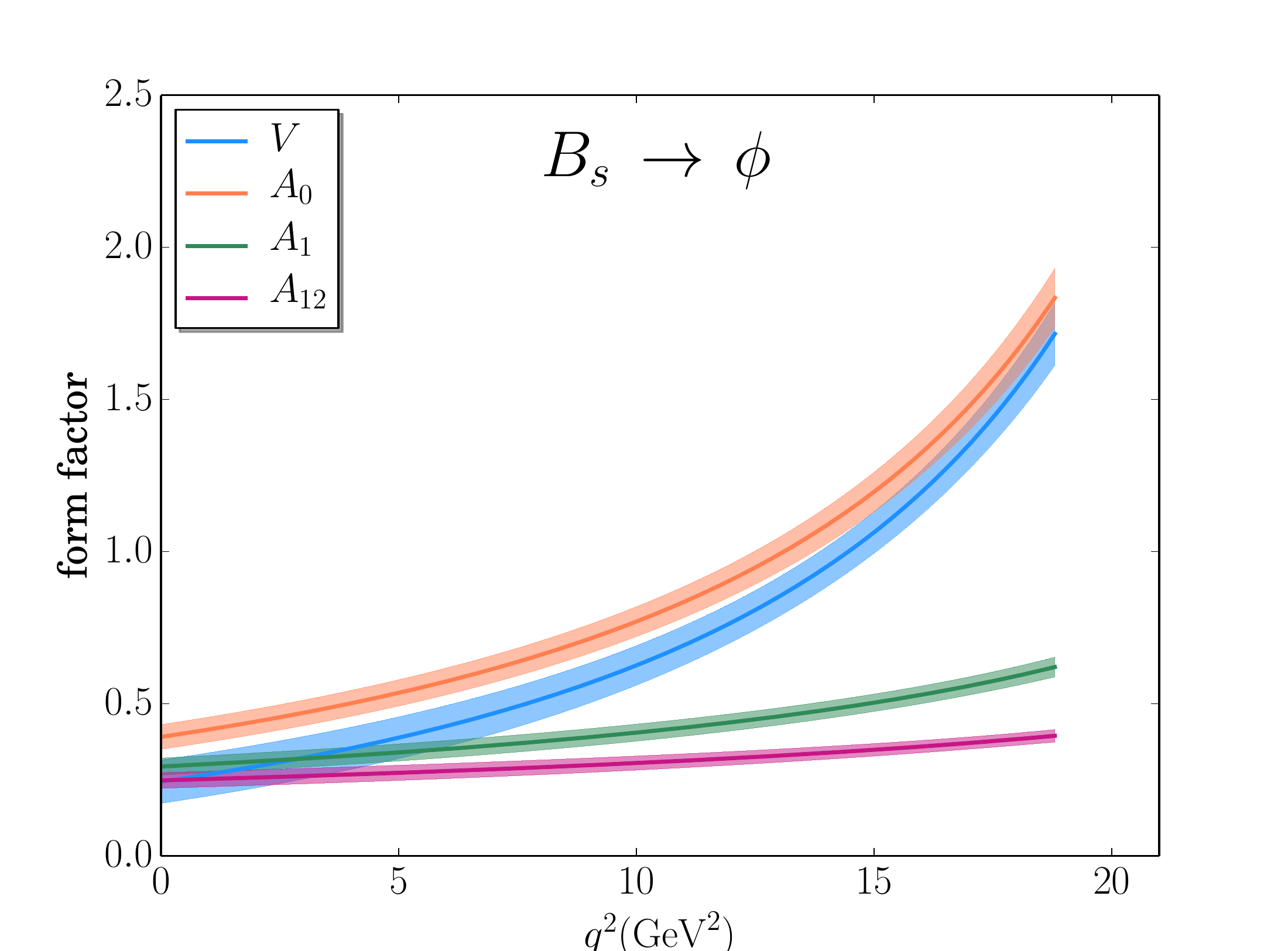}
\includegraphics[width=0.329\textwidth]{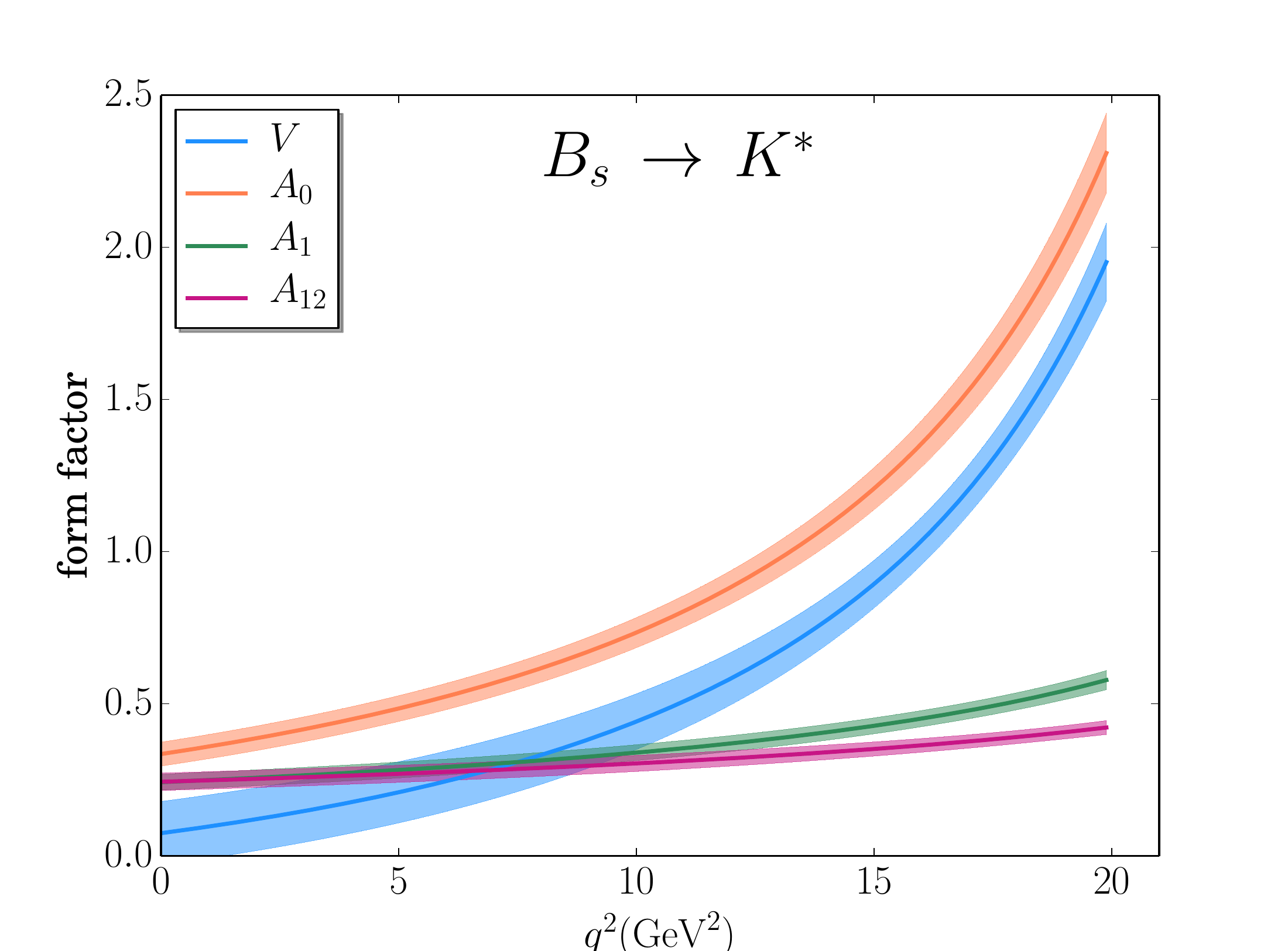}
\includegraphics[width=0.329\textwidth]{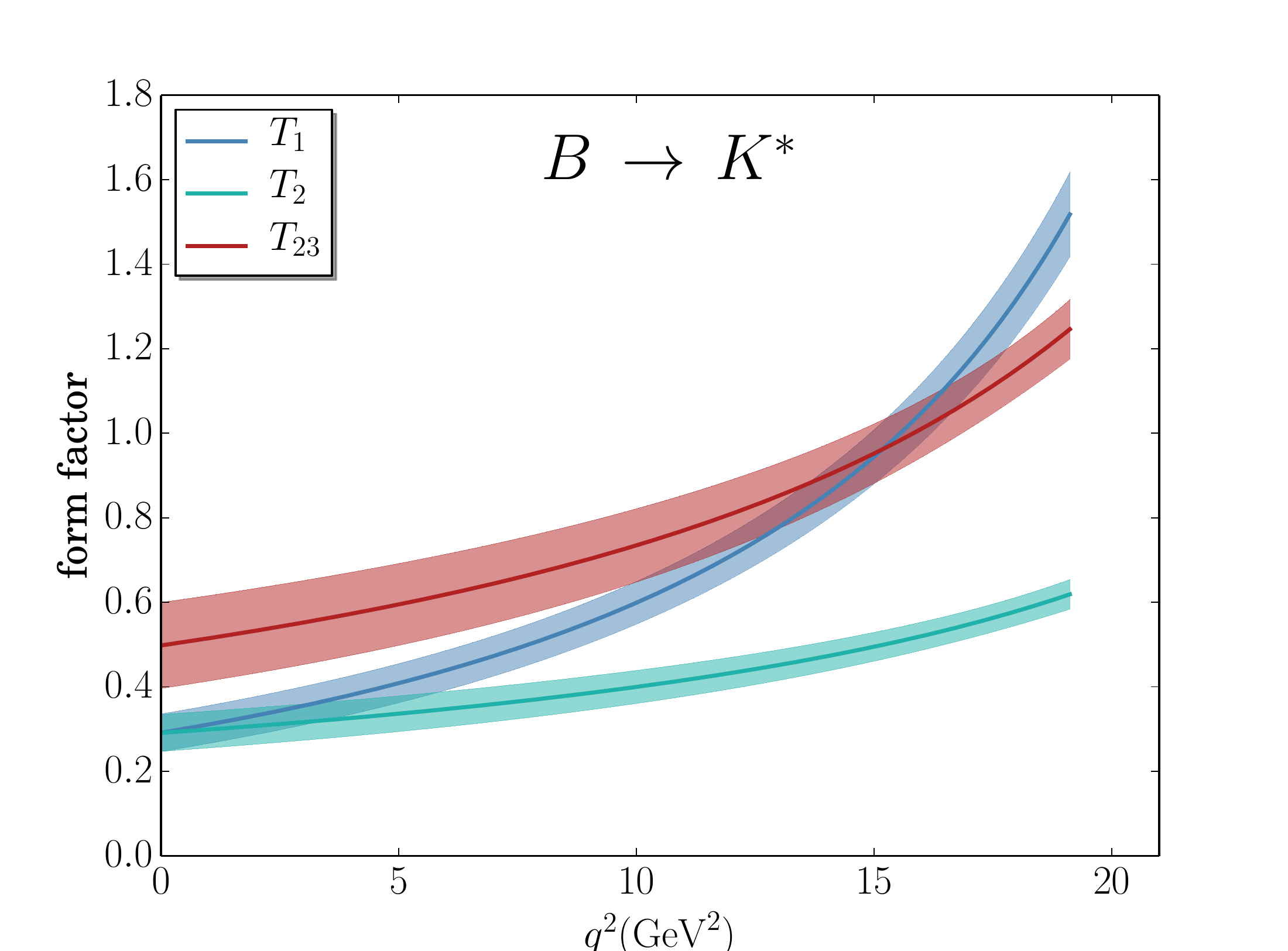}
\includegraphics[width=0.329\textwidth]{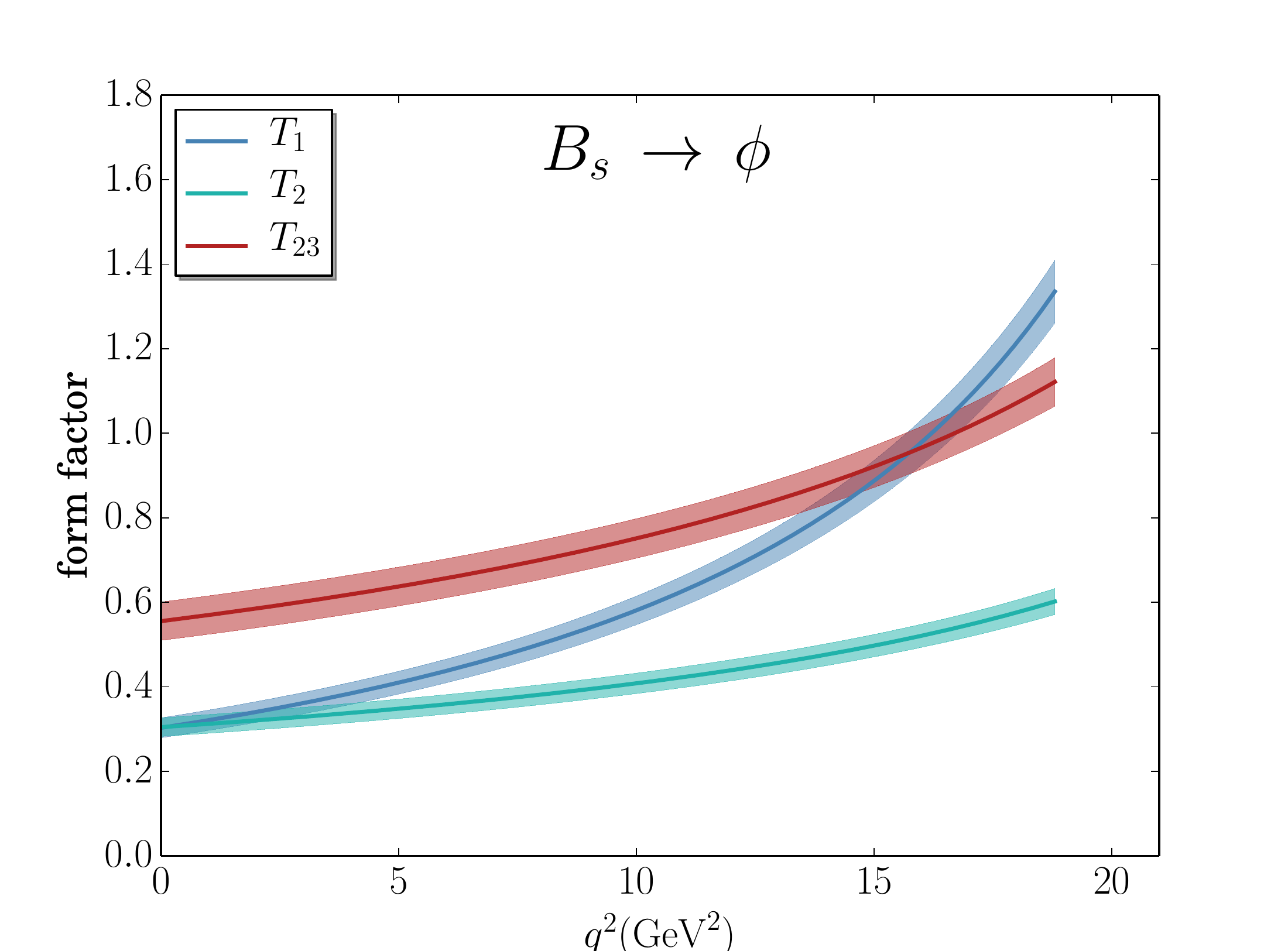}
\includegraphics[width=0.329\textwidth]{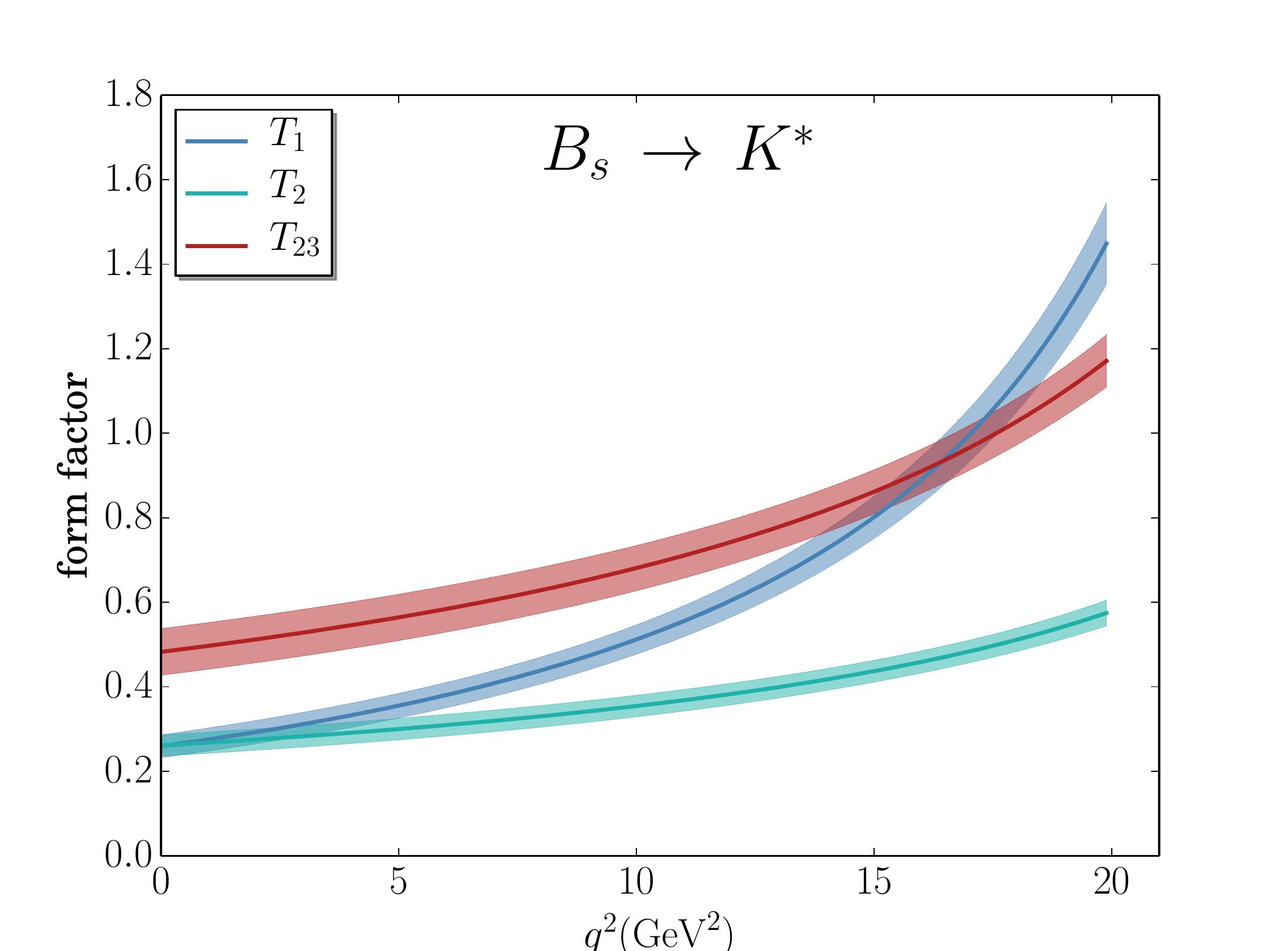}
\caption{\label{fig:FFplots}Lattice QCD determinations of $B \to K^*$ (left),
  $B_s\to \phi$ (middle), and $B_s \to K^*$ (right) form factors in the
  physical limit. Error bands include statistical and systematic uncertainties.}
\end{figure}

At two kinematic points the form factors are not all linearly independent.
When $q^2=0$, equations of motion may be used to relate matrix elements, giving
 two constraints on the form factors. At the kinematic endpoint,
$q^2 = t_- = (m_B - m_V)^2$, two pairs of form factors become linearly 
dependent:
\begin{align}
A_{12}(0) & = \frac{m_B^2 - m_V^2}{8m_B m_V} A_0(0) &
A_{12}(t_-) & = \frac{(m_B + m_V)(m_B^2 - m_V^2 - t_-)}{16m_B m_V^2} A_1(t_-) 
\nonumber \\
T_1(0) &= T_2(0) &
T_{23}(t_-) & = \frac{(m_B + m_V)(m_B^2 +3m_V^2 - t_-)}{8m_B m_V^2}T_2(t_-)\,.
\label{eq:kinconst}
\end{align}
In our published work fits were done to individual form factors separately,
with the exception of a joint fit to $T_1$ and $T_2$, so that the constraint
$T_1(0)= T_2(0)$ could be implemented.  This was the most important constraint
for existing measurements such as $\mathcal{B}(B \to K^*\gamma)$.  In these
proceedings we implement all constraints in (\ref{eq:kinconst}) by performing
simultaneous fits to the vector and axial vector form factors $\{V, A_0, A_1,
A_{12}\}$ and to the tensor and pseudotensor form factors $\{T_1, T_2,
T_{23}\}$.  This will ensure that endpoint relations such as $F_L = \tfrac13$
are precisely satisfied \cite{Hiller:2013cza}.  The constraints are
implemented by adding a fake ``data point'' to the fit requiring that the
left-hand and right-hand sides (in the physical mass limit) are equal up to
some uncertainty.  For $q^2=0$ this uncertainty is set to be $1\%$, an
estimate for the size of $O(z^2)$ terms truncated from our fit ansatz.  For
$q^2 = q^2_{\mathrm{max}}$ we take the uncertainty to be $10^{-4}$ since we
have data in the high-$q^2$ range.  In practice, results do not change if
either of these is reduced.

We have tried including cross-correlations between all of the form factor data
in order to estimate the correlations between the full set of fit parameters.
The data covariance matrix was not determined well enough to allow for a
single fit, so several fits to sets of 3 or 4 form factors were done.  We
suspect that the data covariance matrices were still not entirely
well-determined.  For example, when we included cross-correlations but no new
kinematic constraints compared to our published fits, we found deviations of
approximately 1 standard deviation, despite the fact that the fit parameters
were independent of each other.  Therefore, for our final fit we include only
the data correlations between $A_0$ and $A_{12}$ and between $T_1$ and $T_2$.
These are the pairs of form factors which must satisfy kinematic constraints
(\ref{eq:kinconst}) at $q^2=0$.  Figure~\ref{fig:FFplots} shows the results
for the form factors in the physical limit.  Further details and results of
these fits are appended to the end of the arXiv version of this write-up
[arXiv:1501.00367].

\begin{figure}
\centering
\includegraphics[width=0.32\textwidth]{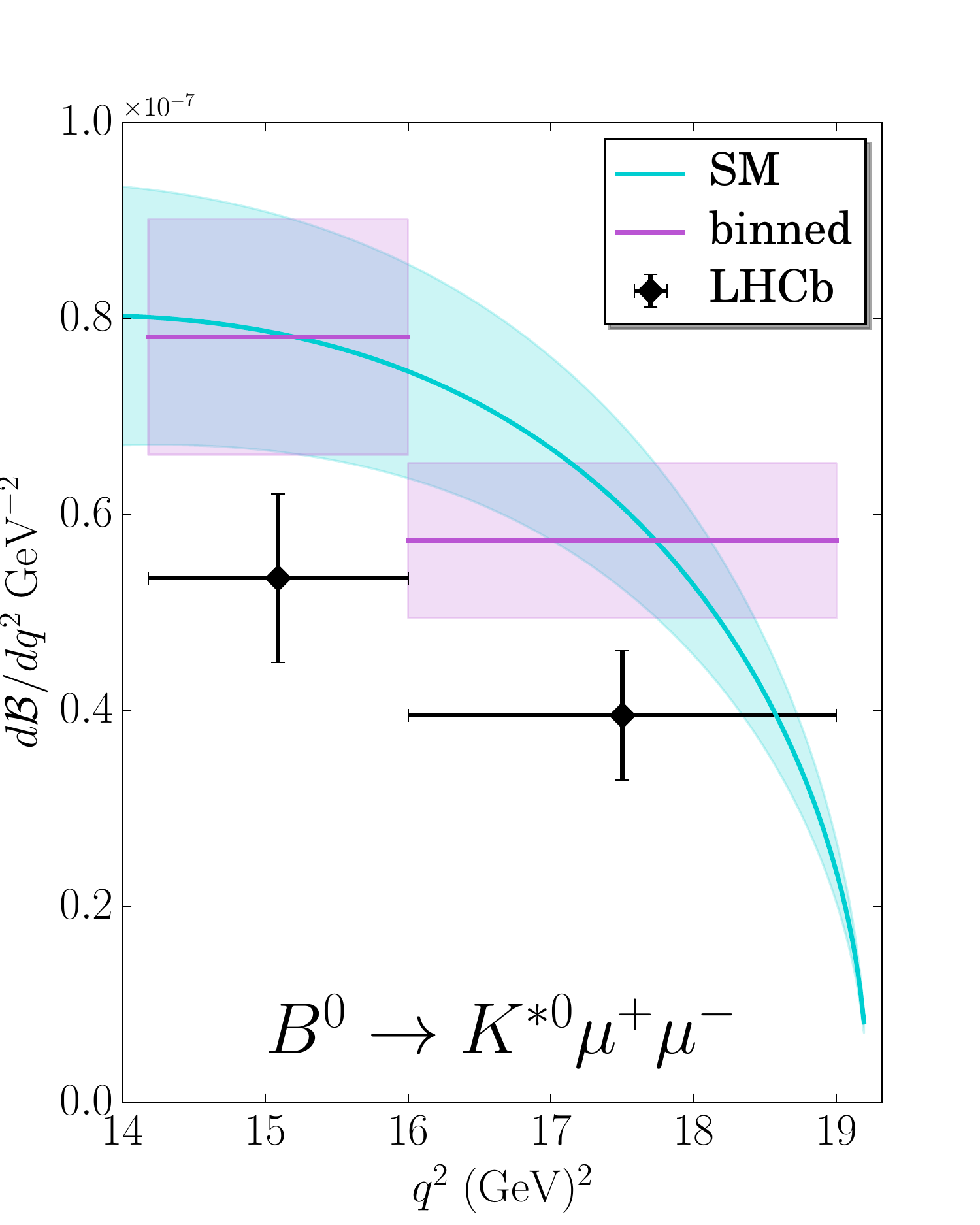}
\includegraphics[width=0.32\textwidth]{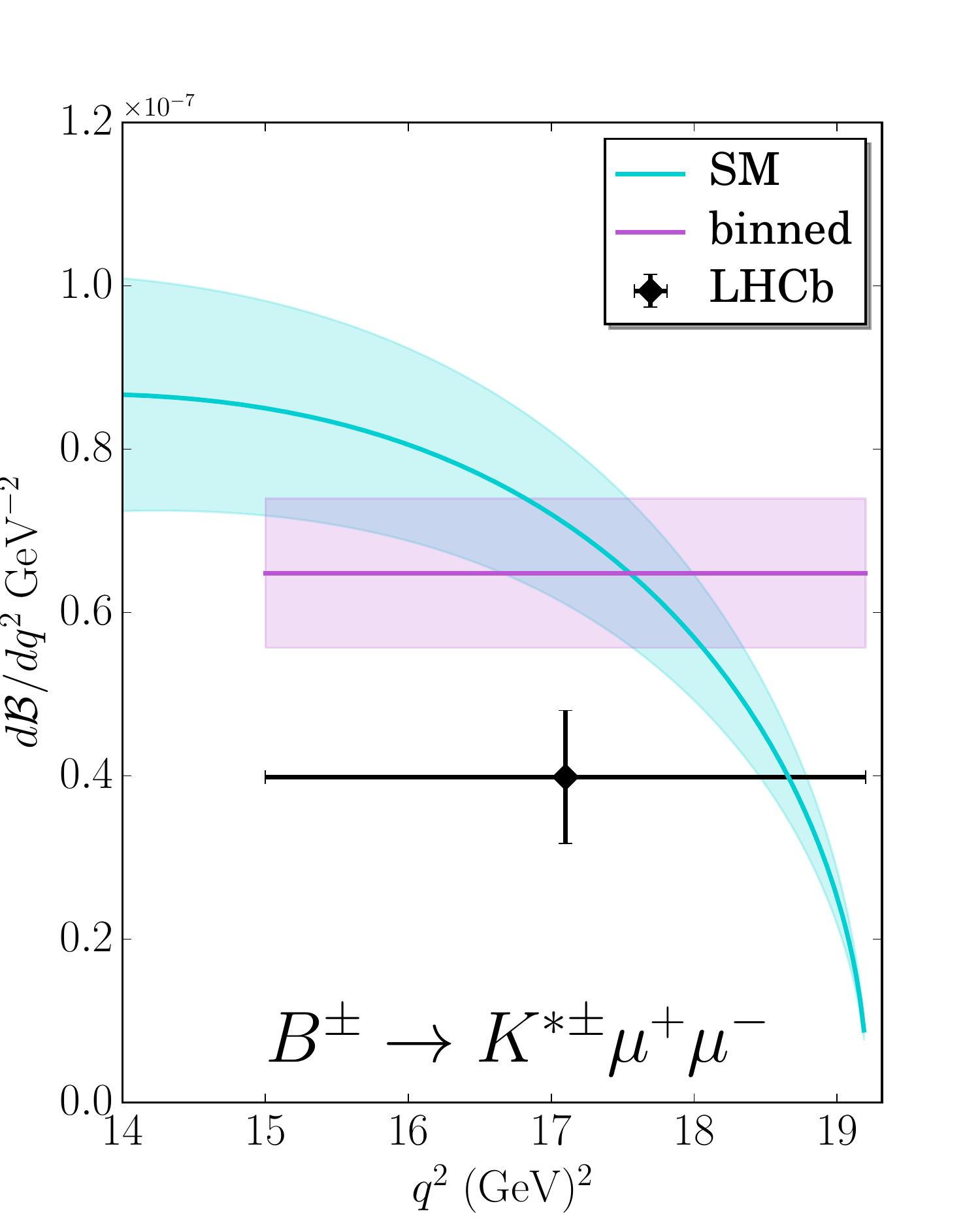}
\includegraphics[width=0.32\textwidth]{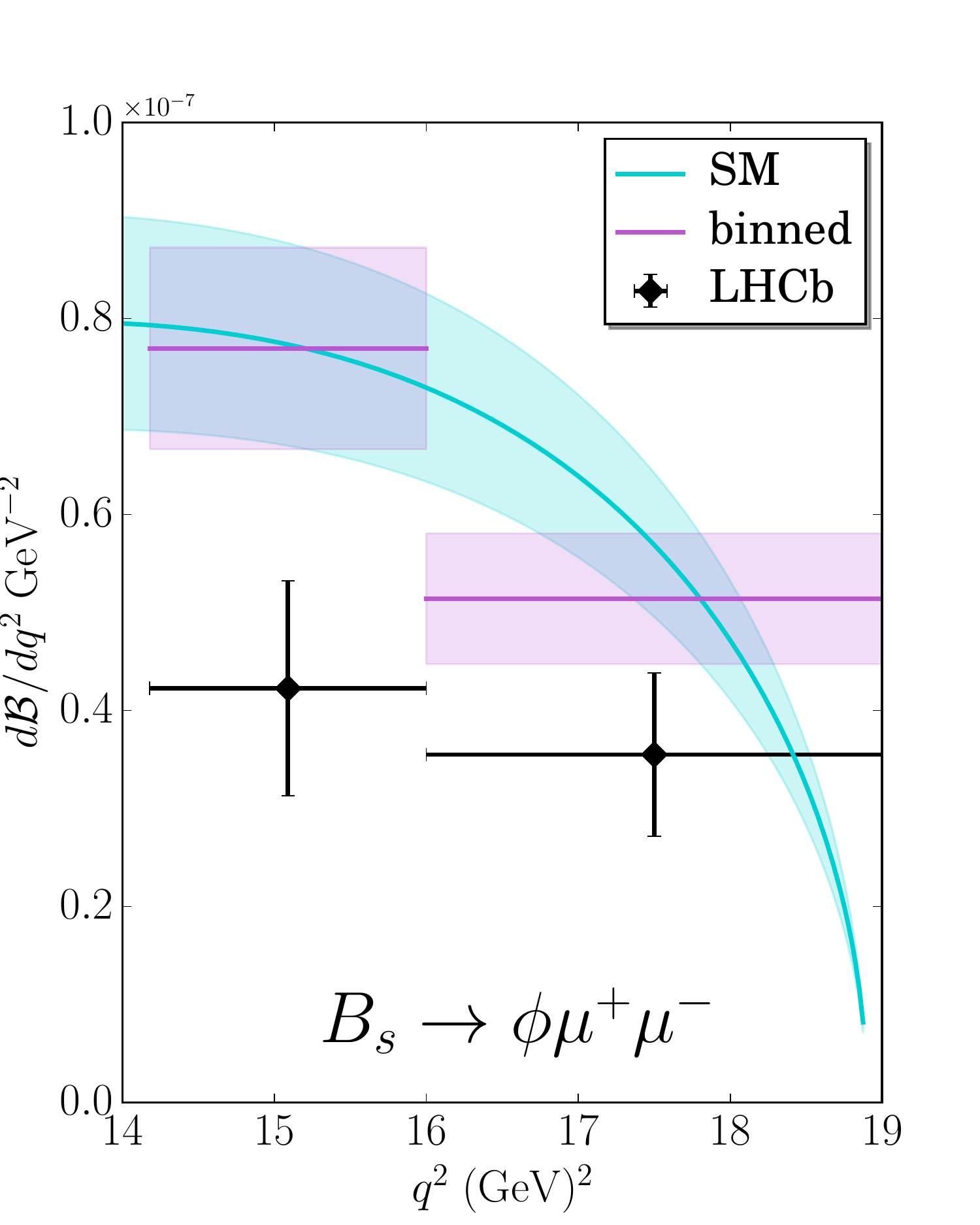}
\caption{\label{fig:bf}Differential branching fractions
for $B^0\to K^{*0}\mu^+\mu^-$ (left), $B^\pm\to K^{*\pm}\mu^+\mu^-$ (middle), 
and $B_s\to \phi\mu^+\mu^-$ (right). Experimental results shown are
from \cite{Aaij:2013iag,Aaij:2014pli,Aaij:2013aln}, respectively.}
\end{figure}

We use these new form factor fits, including the constraints
(\ref{eq:kinconst}) and the estimated correlation matrix, to determine several
$B^0\to K^{*0} \mu^+\mu^-$, $B^\pm\to K^{*\pm} \mu^+\mu^-$, and $B_s\to \phi
\mu^+\mu^-$ observables.  For most quantities, notably the differential
branching fraction, we find only negligible changes to our published results
fits \cite{Horgan:2013pva}.  Fig.~\ref{fig:bf} shows our Standard Model
differential branching fraction compared to experimental data, with the data
consistently lower than the theoretical prediction (see also
Table~\ref{tab:binned_bf}).  Tables~\ref{tab:binned_ang_sl} and
\ref{tab:binned_ang_ss} give Standard Model predictions for angular
observables.  The most significant change compared to \cite{Horgan:2013pva} is
to $F_L$, which shifts and is more precisely determined.  This is due to the
inclusion of the $A_{12}(t_-)/A_1(t_-)$ constraint.  There is also a $\lesssim
1\sigma$ shift in the central value of $S_3$ in the case of $B_s\to \phi
\mu^+\mu^-$.

\begin{figure}
\centering
\includegraphics[width=0.4\textwidth]{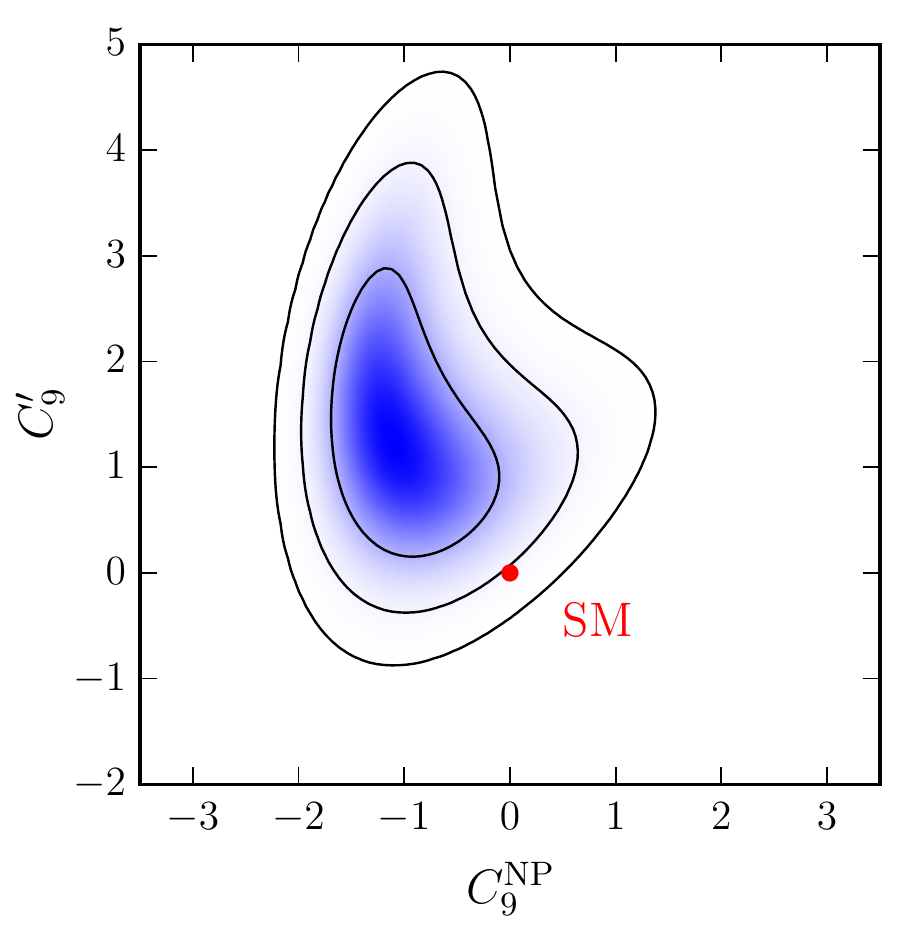}
\caption{\label{fig:likelihood}The likelihood function of a
  2-parameter fit to $B^0 \to K^{*0}\mu^+\mu^-$ and $B_s \to
  \phi\mu^+\mu^-$ experimental data with $q^2 > 14.18$ GeV${}^2$.  The
  Standard Model (SM) value lies just outside the $2\sigma$
  contour.}
\end{figure}

We have repeated our beyond-the-Standard-Model fit to large-$q^2$
experimental data for $B^0 \to K^{*0}\mu^+\mu^-$ and $B_s \to
\phi\mu^+\mu^-$, allowing the Wilson coefficients $C_9 =
C_9^{\mathrm{SM}} + C_9^{\mathrm{NP}}$ and $C_9'$ to deviate from
their Standard Model values ($C_9'{}^{\mathrm{SM}} \approx 0$). This
yields $C_9^{\mathrm{NP}} = -1.1 \pm 0.5$ and $C_9' = 1.2 \pm 0.9$,
comparable to what we found before \cite{Horgan:2013pva}.
Figure~\ref{fig:likelihood} shows a contour plot of the likelihood
function of this fit.

\begin{table}
\centering
\begin{tabular}{cccc} \hline\hline
$q^2$ bin (GeV${}^2$) & {$B^0 \to K^{*0}\mu^+\mu^-$} &
{$B^\pm \to K^{*\pm}\mu^+\mu^-$} &
{$B_s \to \phi\mu^+\mu^-$} \\
\hline
$[14.18, 16.00]$ & $7.8(1.2)\times 10^{-8}$ & $8.4(1.3)\times 10^{-8}$ 
& $7.7(1.0) \times 10^{-8}$\\
$[16.00, 19.00]$ & $5.73(79)\times 10^{-8}$   & $6.19(85) \times 10^{-8}$
& $5.14(67)\times 10^{-8}$ \\
$[14.18, 19.00]$ & $6.52(94)\times 10^{-8}$   & $7.0(1.0)\times 10^{-8}$ 
& $6.11(80)\times 10^{-8}$ \\
\hline\hline
\end{tabular}
\caption{\label{tab:binned_bf}Standard model predictions for branching
fractions $d\mathcal{B}/dq^2 (\mathrm{GeV}^{-2})$ in bins of $q^2$.}
\end{table}

\begin{table}
\centering
\begin{tabular}{cccccc} \hline\hline
$q^2$ bin (GeV${}^2$) & $F_L$ & $A_{FB}$ & $S_3$ & $P_4'$ & $P_5'$ \\ \hline
$[14.18, 16.00]$ & 0.360(42) & 0.410(35) & $-0.160(29)$ & 0.612(17) & 
$-0.702(59)$ \\
$[16.00, 19.00]$ & 0.336(25) & 0.347(21) & $-0.230(17)$ & 0.650(08) &
 $-0.541(35)$ \\
$[14.18, 19.00]$ & 0.347(32) & 0.375(26) & $-0.198(22)$ & 0.633(12) & 
$-0.614(45)$ \\
\hline\hline
\end{tabular}
\caption{\label{tab:binned_ang_sl}Standard model predictions for 
$B^0 \to K^{*0}\mu^+\mu^-$ angular observables in bins of $q^2$.}
\end{table}

\begin{table}
\centering
\begin{tabular}{cccc} \hline\hline
$q^2$ bin (GeV${}^2$) & $F_L$ & $S_3$ & $P_4'$  \\ \hline
$[14.18, 16.00]$ & 0.382(20) & $-0.172(13)$ & 0.624(8) \\
$[16.00, 19.00]$ & 0.347(12) & $-0.242(08)$ & 0.659(4) \\
$[14.18, 19.00]$ & 0.364(15) & $-0.209(10)$ & 0.642(5) \\
\hline\hline
\end{tabular}
\caption{\label{tab:binned_ang_ss}Standard model predictions for 
$B_s \to \phi\mu^+\mu^-$ angular observables in bins of $q^2$.}
\end{table}

\section{Open issues}
\label{sec:open}

One uncontrolled approximation in our calculation is that we fit correlation
functions using a single interpolating operator for the vector meson final
state, assuming that it corresponds to the $K^*$ or $\phi$.  A fully
controlled calculation would include scattering states in the analysis. In
order to do so, a much more expensive and complicated set of calculations must
be undertaken. The path forward was set out in a paper which appeared during
this conference \cite{Briceno:2014uqa}.  Work to study the spectrum of $K\pi$
and $K\eta$ states has recently begun
\cite{Prelovsek:2013ela,Dudek:2014qha,Wilson:2014cna}, marking the first major
step toward a full lattice calculation of $B \to K^*(\to K\pi)$ matrix
elements.  In the interim, the use of our form factor results comes with the
assumption that threshold effects will be small.  One might expect them to be
smaller for $B_s \to \phi$ than for $B\to K^*$ since the $\phi$ is relatively
narrow.  One might also note that heavy meson chiral perturbation theory
predicts percent-level threshold effects in $B\to D^*$ form factors
\cite{Randall:1993qg,Hashimoto:2001nb} and hope that such is the case for
lighter-mass mesons.  Nevertheless at least there is a plan to systematically
include this effect in the future.

Perhaps a larger open question regards the extent to which form
factors are sufficient to determine the hadronic contributions to
observables.  In addition to matrix elements of 2-quark operators
(\ref{eq:ops}), there are matrix elements of non-local operators
$O_i(x) j_\mu(y)$, notably where $O_i$ is a 4-quark $b\to s$ operator
which creates a $\bar{c}c$ pair, annihilated at another point
by the vector current $j_\mu$.  It was expected that these matrix
elements would make theoretical predictions unreliable for $\sqrt{q^2}
\approx m_{J/\psi}$ or $m_{\psi'}$.  Within the context of an operator
product expansion, leading-order contributions from these non-local
operators can be (and, in our work, have been) included
\cite{Grinstein:2004vb,Beylich:2011aq}.  However, contributions from
the $\psi(4160)$ to the $B^+\to K^+\mu^+ \mu^-$ are larger than
anticipated \cite{Aaij:2013pta} and the same appears to be true for
decays to vector mesons.  Understanding nonfactorizable contributions
is probably the most important open issue which theoretical
predictions must confront \cite{Lyon:2014hpa}.

\section{Conclusions}
\label{sec:concl}

This write-up summarizes our recent calculation of $B\to K^*$, $B_s\to \phi$,
and $B_s \to K^*$ form factors.  Here we give a slightly improved set of fits
to the form factor shapes, including all four kinematic constraints and
estimates of the more significant correlations between form factor parameters.
Given that this update is only a minor improvement upon our published work and
has a small effect on observables, we would be grateful if those authors using
the form factor results presented here would cite Ref.~\cite{Horgan:2013hoa}
in addition to this work.

\section*{Acknowledgments}

We are grateful to R.~Zwicky for discussions about the form factor
fits and to the MILC Collaboration for making their gauge-field
configurations publicly available.  This work was supported in
part by the STFC (UK), the DoE (US), and NSFC (China).

\bibliographystyle{../../../Notes_mbw/Bibtex/h-physrev5}
\bibliography{../../../Notes_mbw/Bibtex/mbw}

\newpage
\appendix
\section{Supplemental material}

In this Appendix we give detailed tables in order for readers to use
our new fits to reconstruct the form factors (in the physical limit).
We also compare the new fits to the published ones \cite{Horgan:2013hoa}.

Table~\ref{tab:dmres} gives the mass differences between the $B$ or
$B_s$ and the resonance (or effective pole) relevant for the various
form factors.  These enter the prefactor of the form factor shape
(\ref{eq:sse_3par}).  The remaining tables give results for the
constant and linear coefficients ($a_0$ and $a_1$) of $z$ in the form
factor shape (\ref{eq:sse_3par}): Tables~\ref{tab:av_sl} and
\ref{tab:t_sl} for $B \to K^*$ decays and Tables~\ref{tab:av_ss} and
\ref{tab:t_ss} for $B_s\to \phi$ decays.  In Tables~\ref{tab:av_ls}
and \ref{tab:t_ls} we give results for $B_s \to K^*$ decays, such as
the $b\to u$ decay $B_s \to K^* \ell\nu$ and the $b\to d$ decay
$B_s\to K^*\ell\ell$.  These fit results come from separate fits for
each Table, with correlations included only between $A_0$ and
$A_{12}$ and between $T_1$ and $T_2$.  Other data correlations were
not included for reasons discussed in
Sec.~\ref{sec:fits}.\footnote{ArXiv version 1 of this write-up gave an
  estimate of the full correlation matrix using fits to subsets of
  form factors.  However, the resulting matrix was not positive
  semi-definite. One could devise a method for removing the negative
  eigenvalues, but since the differences between the fits given in
  versions 1 and 2 of these proceedings are not significant compared
  to other uncertainties, we believe the neglected correlations are
  presently insignificant.}

Those wishing to use our form factor calculations for predictions of
observables need not include all the correlation data.  Large $q^2$
observables predominantly depend on $V$ and $A_1$, and their fit
parameters are not very correlated.  On the other hand, the parameters
for individual form factors, $a_0$ and $a_1$, are highly correlated,
so the uncertainty of the LQCD-determined form factor $F(t)$
[Eq.~(\ref{eq:sse_3par})]
is given (in the physical limit) by 
\begin{equation}
\delta F(t) = \frac{1}{1 - t/(m_{B_{(s)}} + \Delta m^F)^2}
\left[(\delta a_0^F)^2 \,+\, (\delta a_1^F)^2 z(t;t_0)^2
\,+\, 2 C_{a_0^F,a_1^F} (\delta a_0^F)(\delta a_1^F)\,z(t;t_0)\right]^\frac{1}{2}
\,.
\end{equation}
Since physical results are obtained by setting $\Delta x = \Delta x_s
= 0$ in (\ref{eq:sse_3par}), one does not need our results for
$c_{01}$ or $c_{01s}$ except to look at the quark-mass dependence of
the form factors.  Data files containing the fit results as tabulated in
Tables \ref{tab:av_sl}-\ref{tab:t_ls} are included as ancillary files
with the arXiv submission.

\begin{table}
\centering
\begin{tabular}{cd{0}d{0}d{0}} \hline \hline
Form factor $F$ & \multicolumn{1}{c}{$B \to K^*$} & 
\multicolumn{1}{c}{$B_s \to \phi$} & \multicolumn{1}{c}{$B_s \to K^*$} \\ \hline
$A_0$ &  87 & 0 & -87 \\
$V$, $T_1$ &  135 & 45 & -42 \\
$A_1$, $A_{12}$, $T_2$, $T_{23}$ & 550 & 440 & 350 \\
\hline \hline
\end{tabular}
\caption{\label{tab:dmres} Mass differences $\Delta m^F$ (in MeV), between the
initial state and pertinent resonance (or effective pole) contributing to
form factor $F$.}
\end{table}

\begin{figure}
\begin{center}
\includegraphics[width=0.495\textwidth]{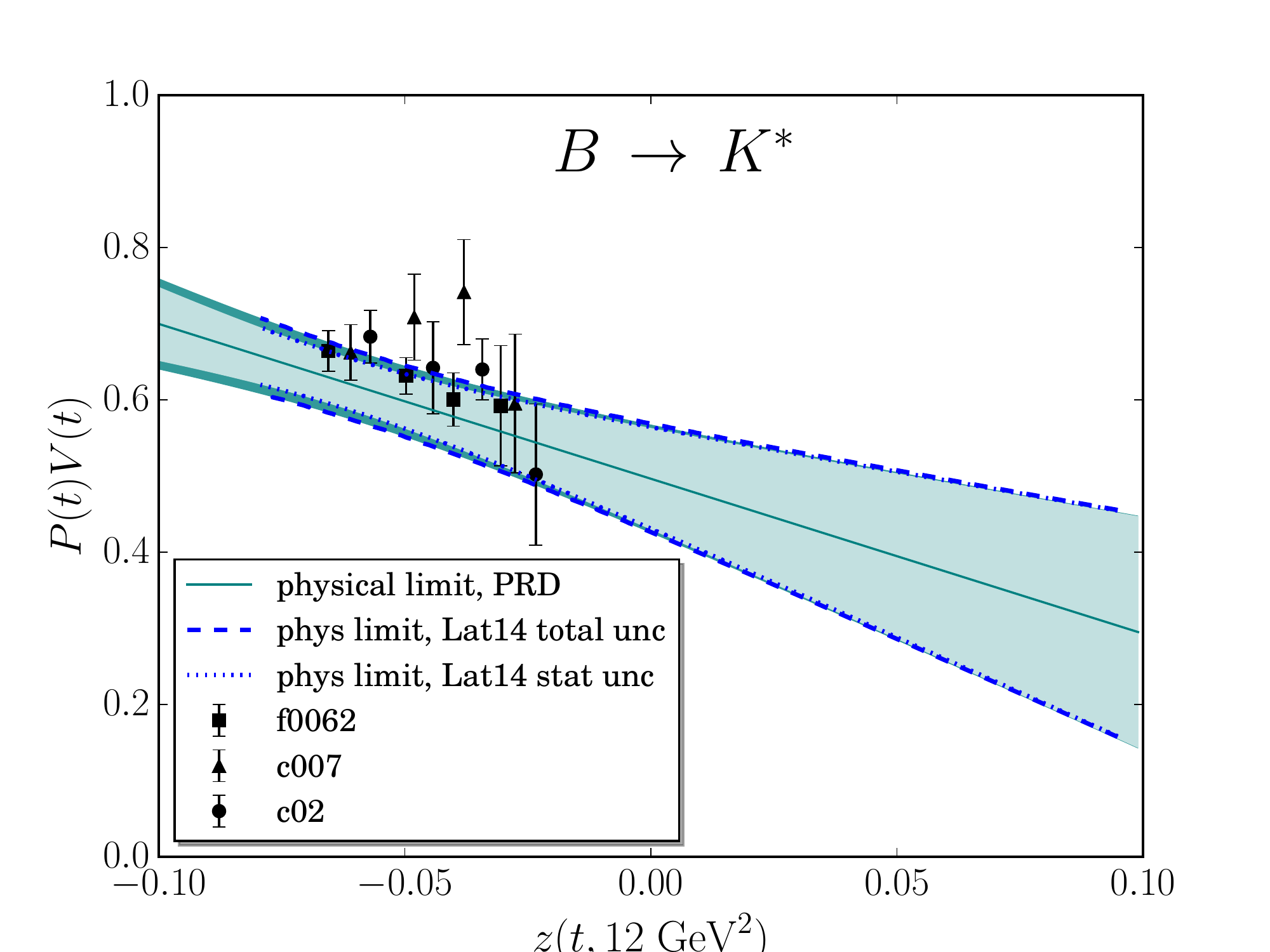}
\includegraphics[width=0.495\textwidth]{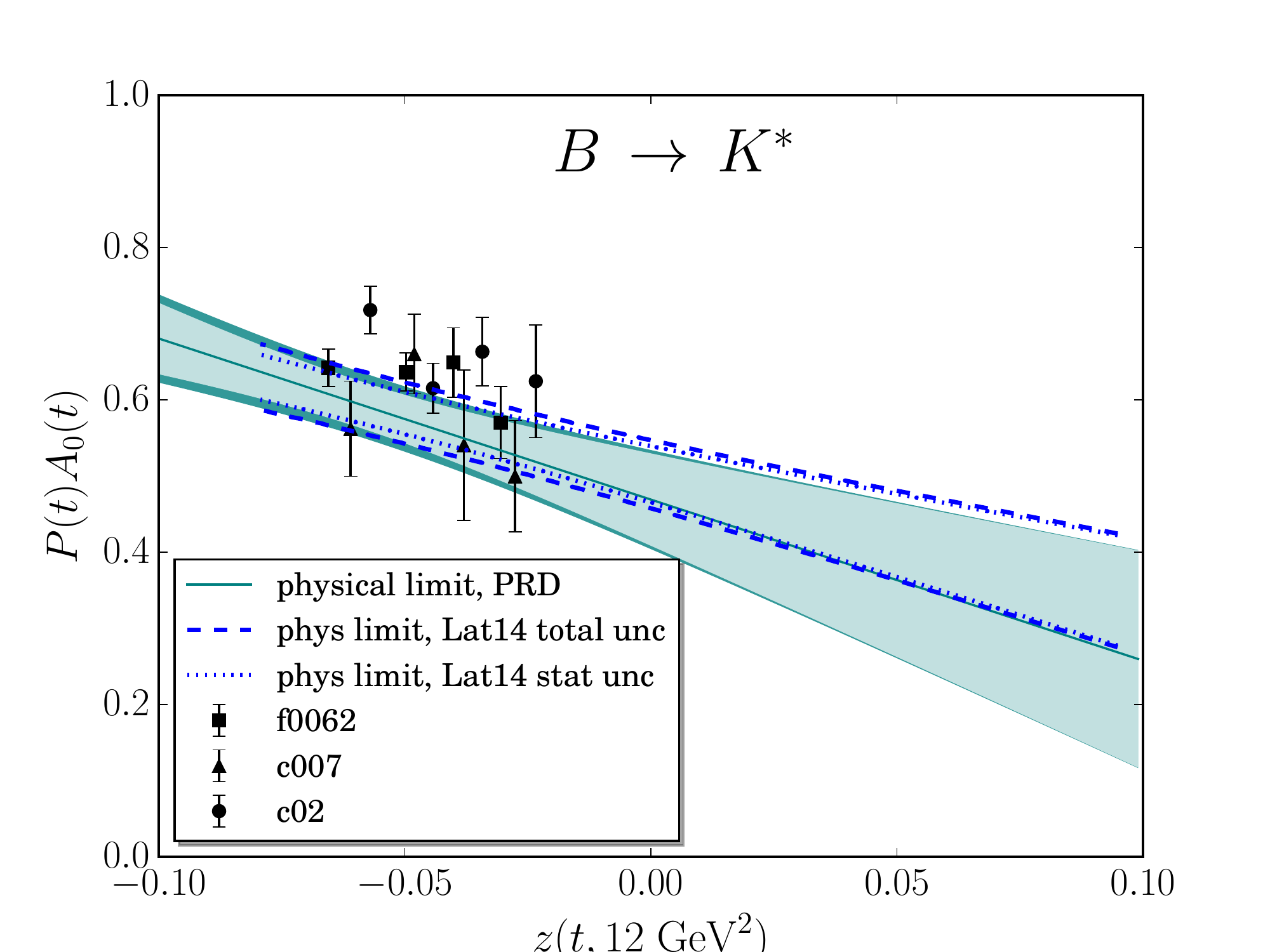}
\includegraphics[width=0.495\textwidth]{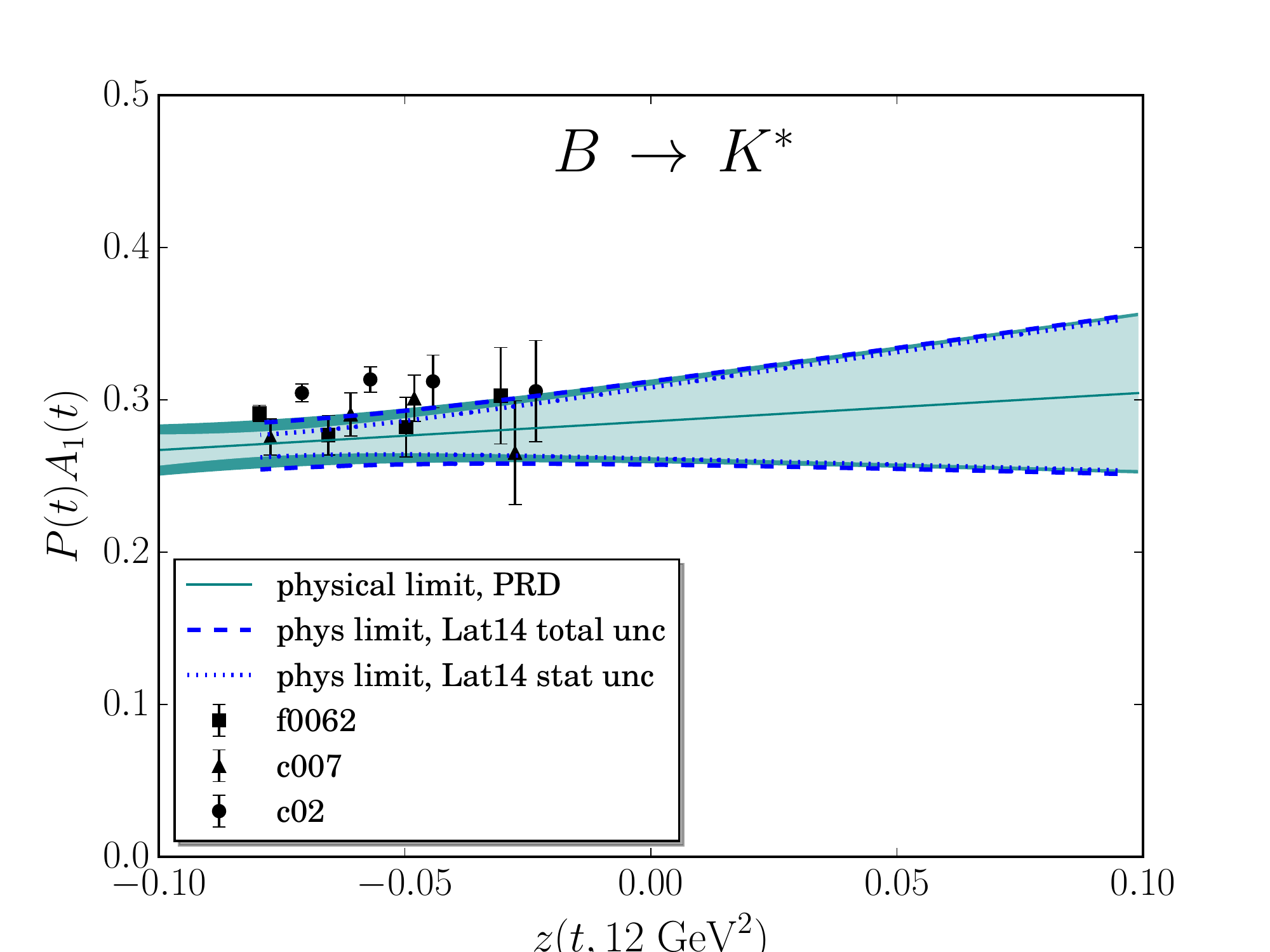}
\includegraphics[width=0.495\textwidth]{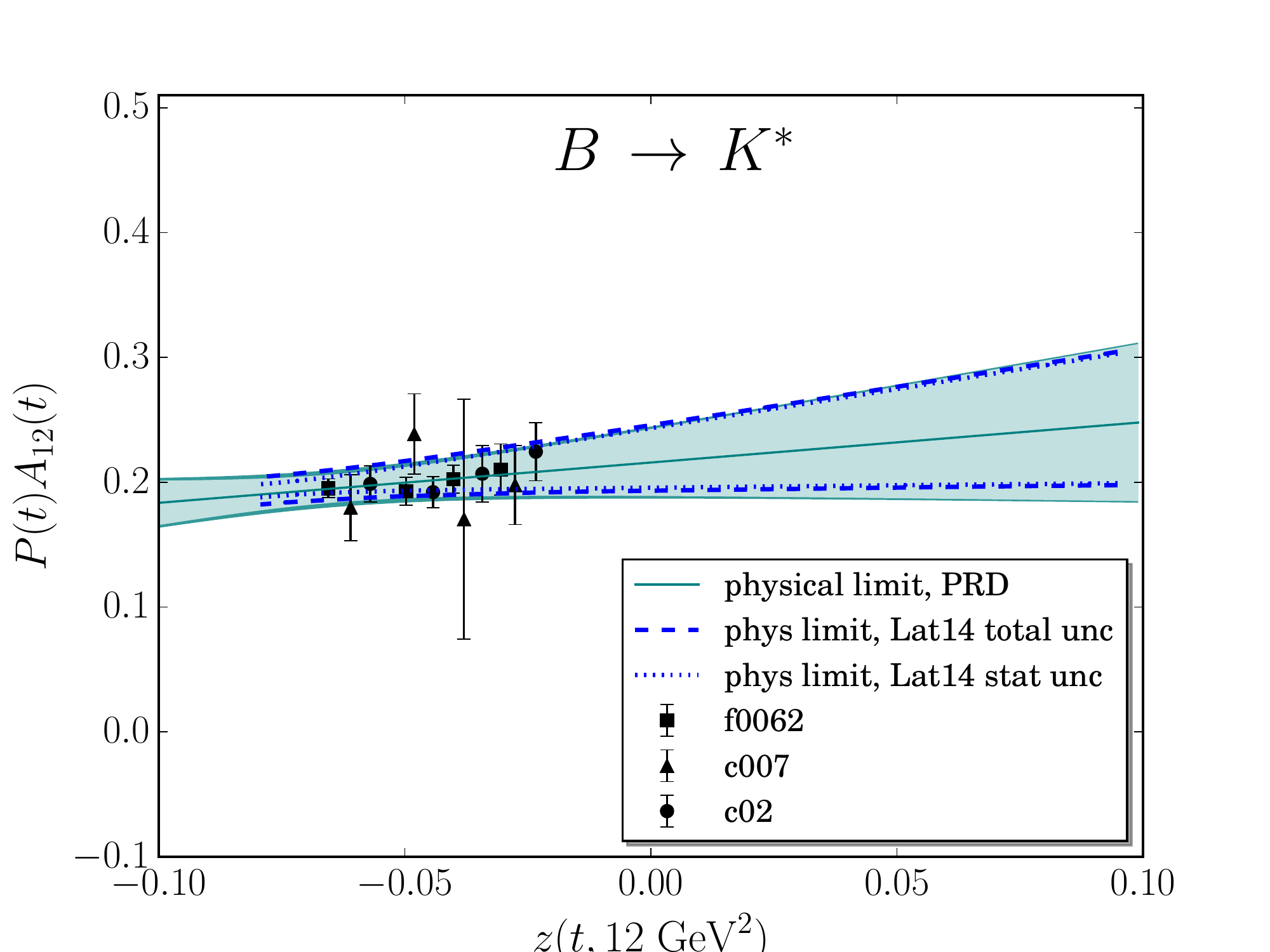}
\includegraphics[width=0.495\textwidth]{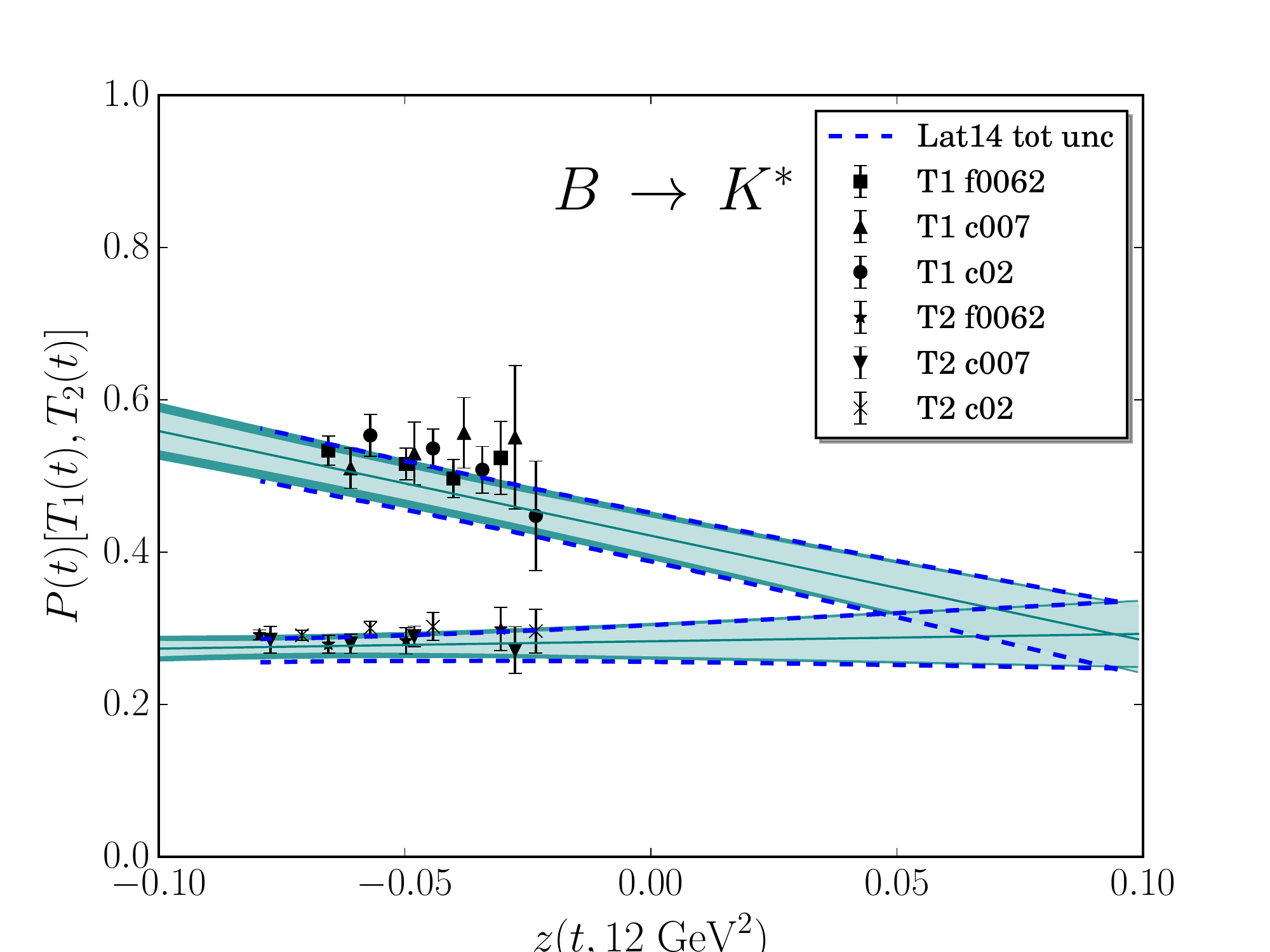}
\includegraphics[width=0.495\textwidth]{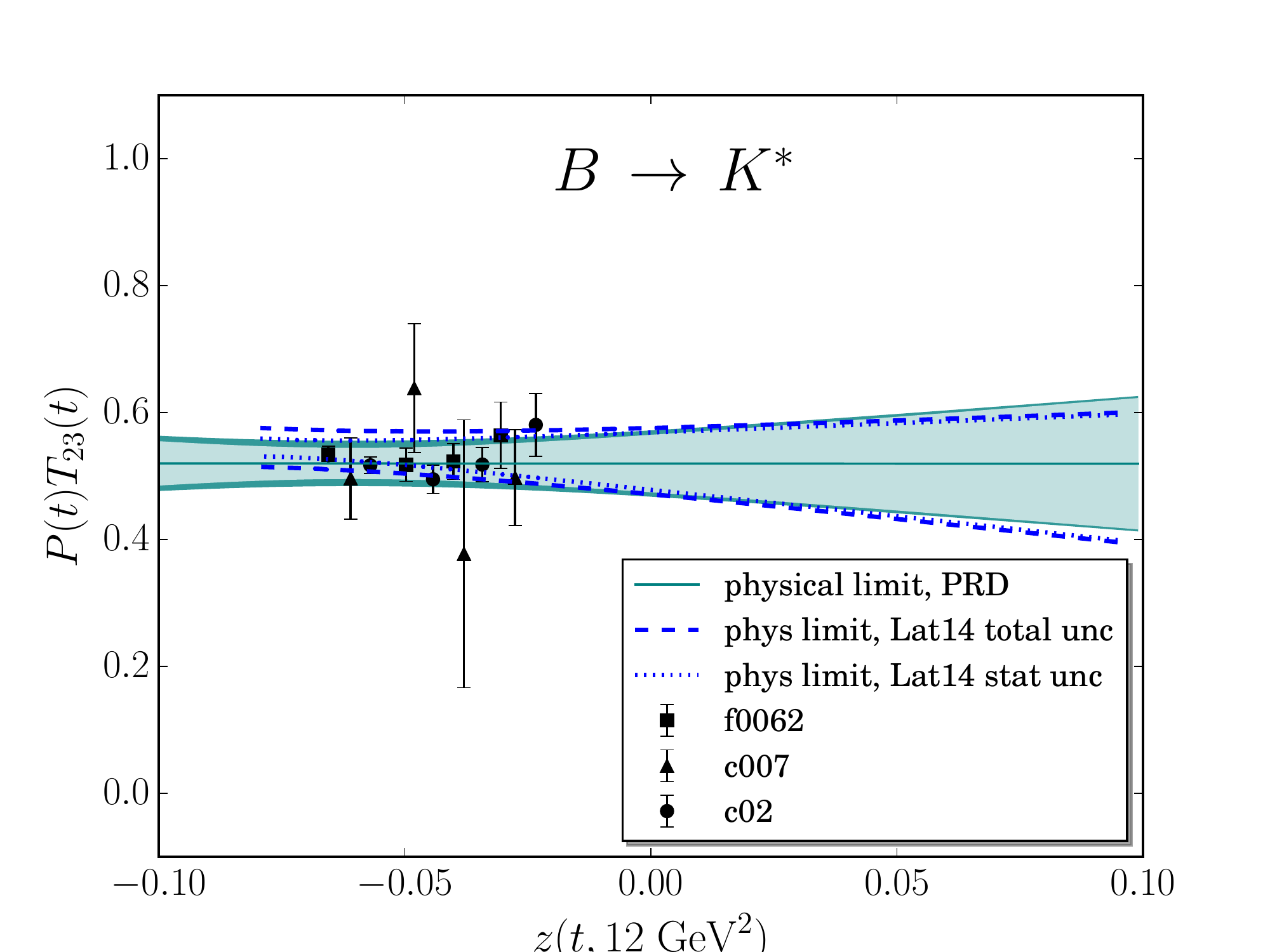}
\end{center}
\caption{\label{fig:pff_sl_hybrid}Comparison of $B\to K^*$ form factor
  fits presented here (blue dashed and dotted lines) compared to
  published fits \cite{Horgan:2013hoa} (teal bands).  Black symbols are
  LQCD results at unphysical quark masses, whereas curves and bands
  are extrapolated to the physical limit.}
\end{figure}

\begin{figure}
\begin{center}
\includegraphics[width=0.495\textwidth]{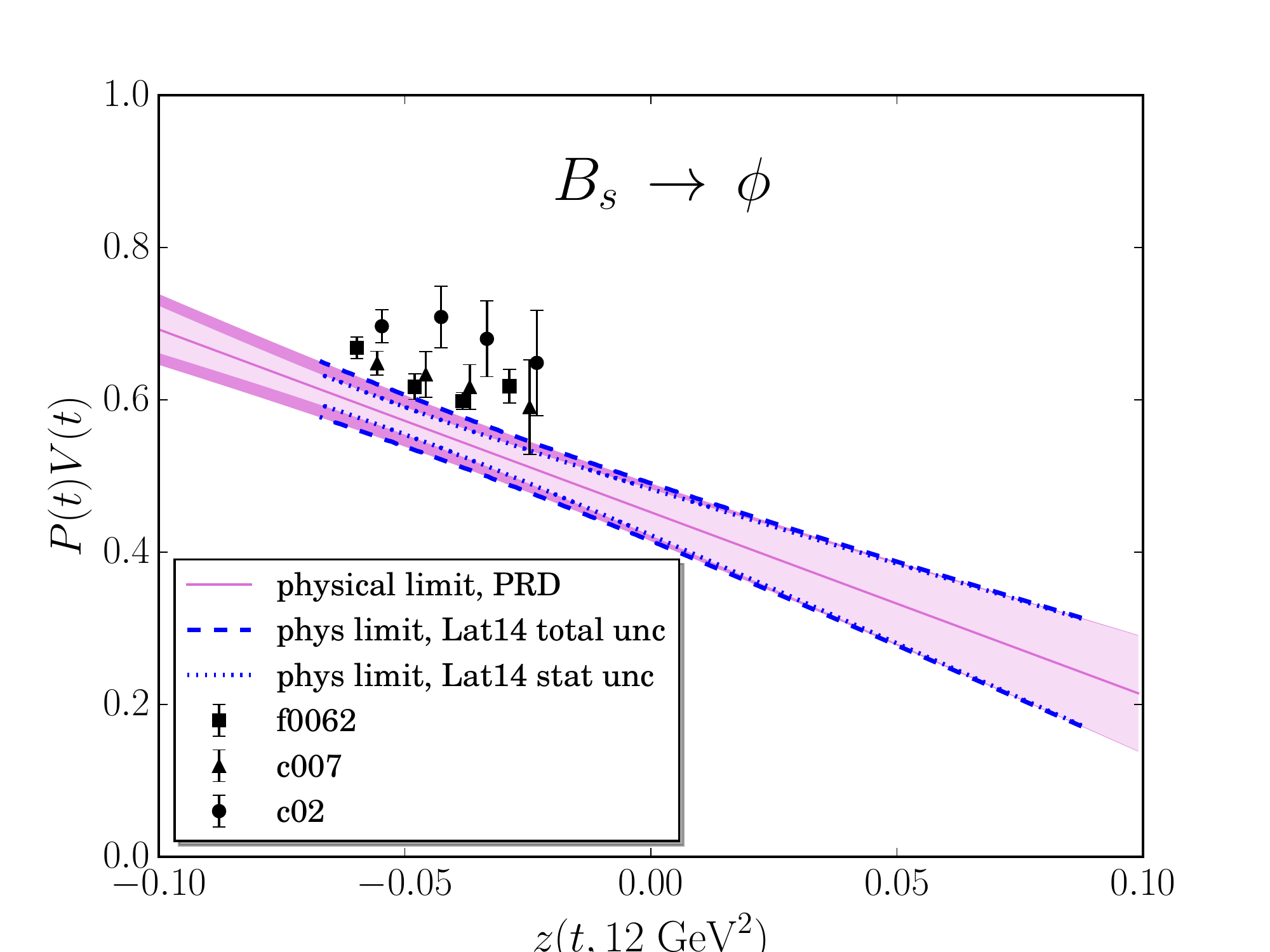}
\includegraphics[width=0.495\textwidth]{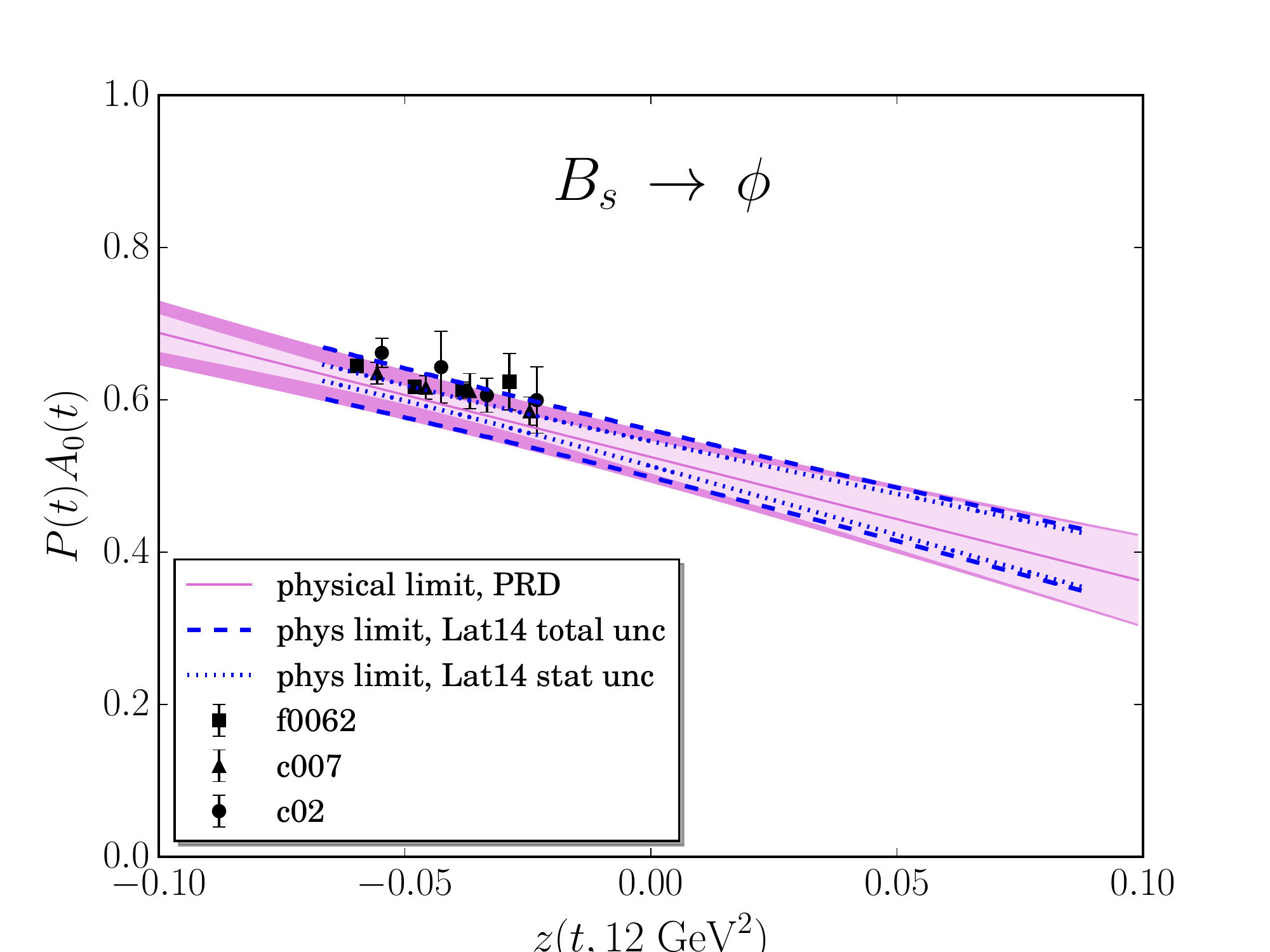}
\includegraphics[width=0.495\textwidth]{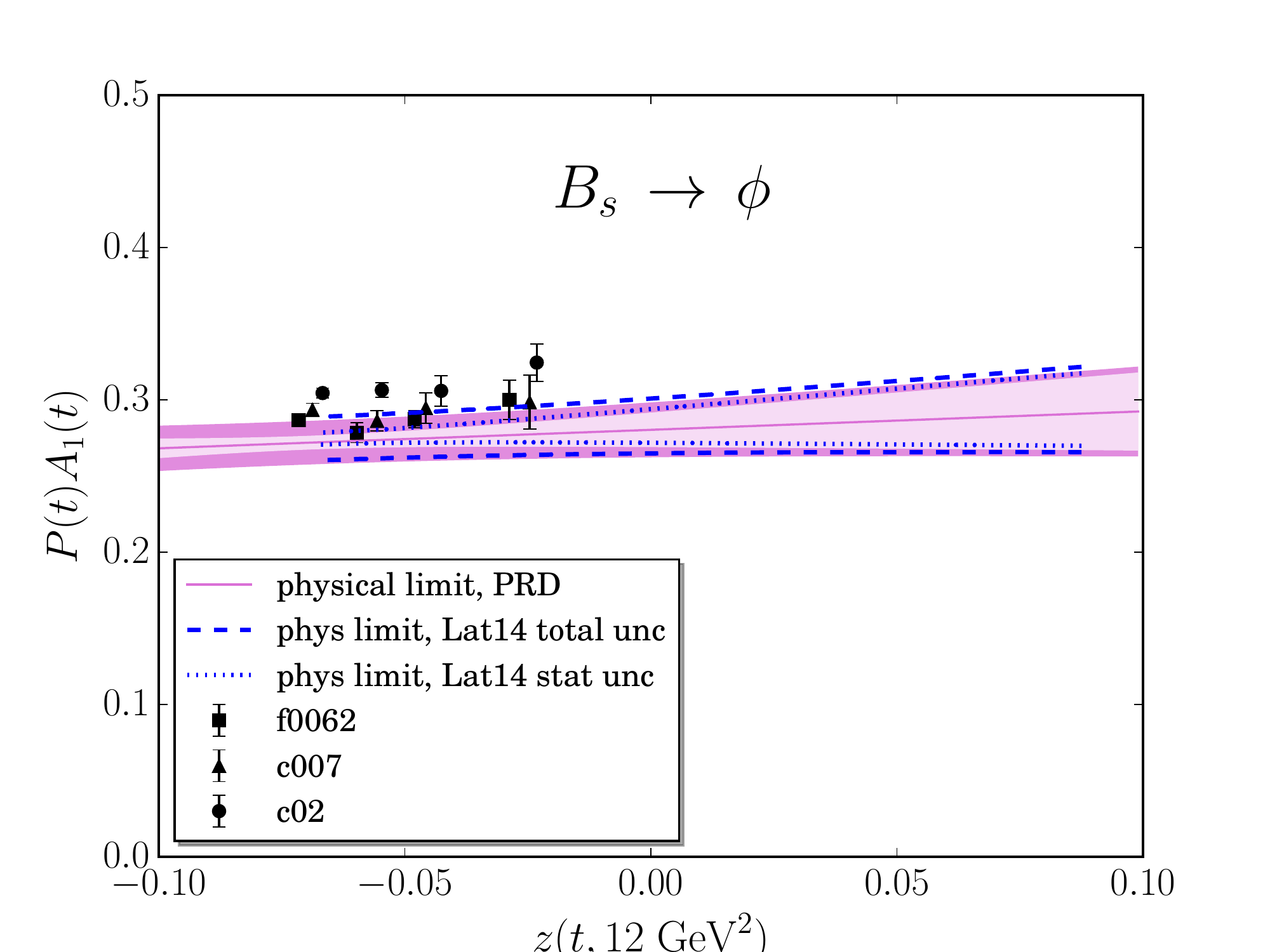}
\includegraphics[width=0.495\textwidth]{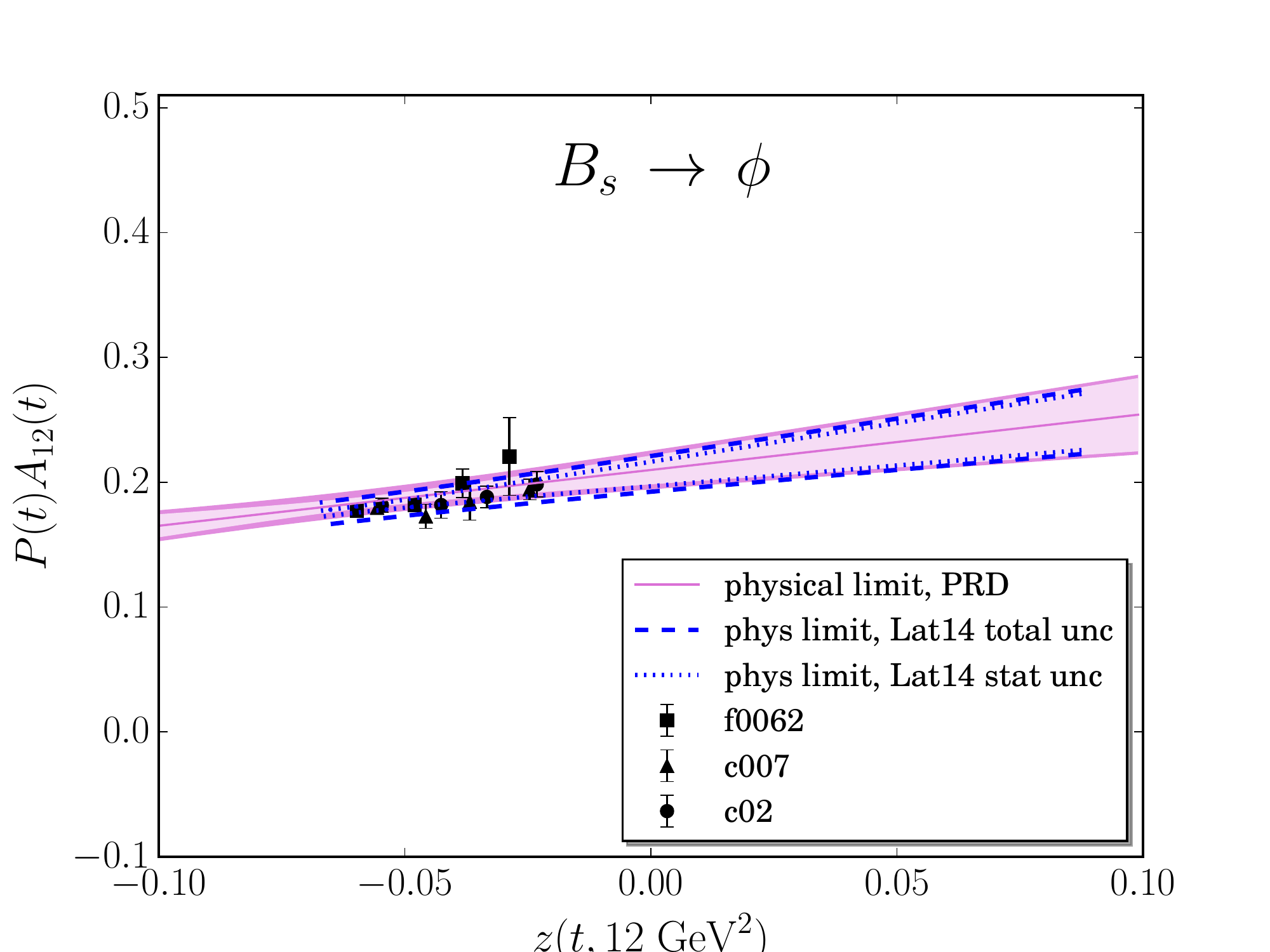}
\includegraphics[width=0.495\textwidth]{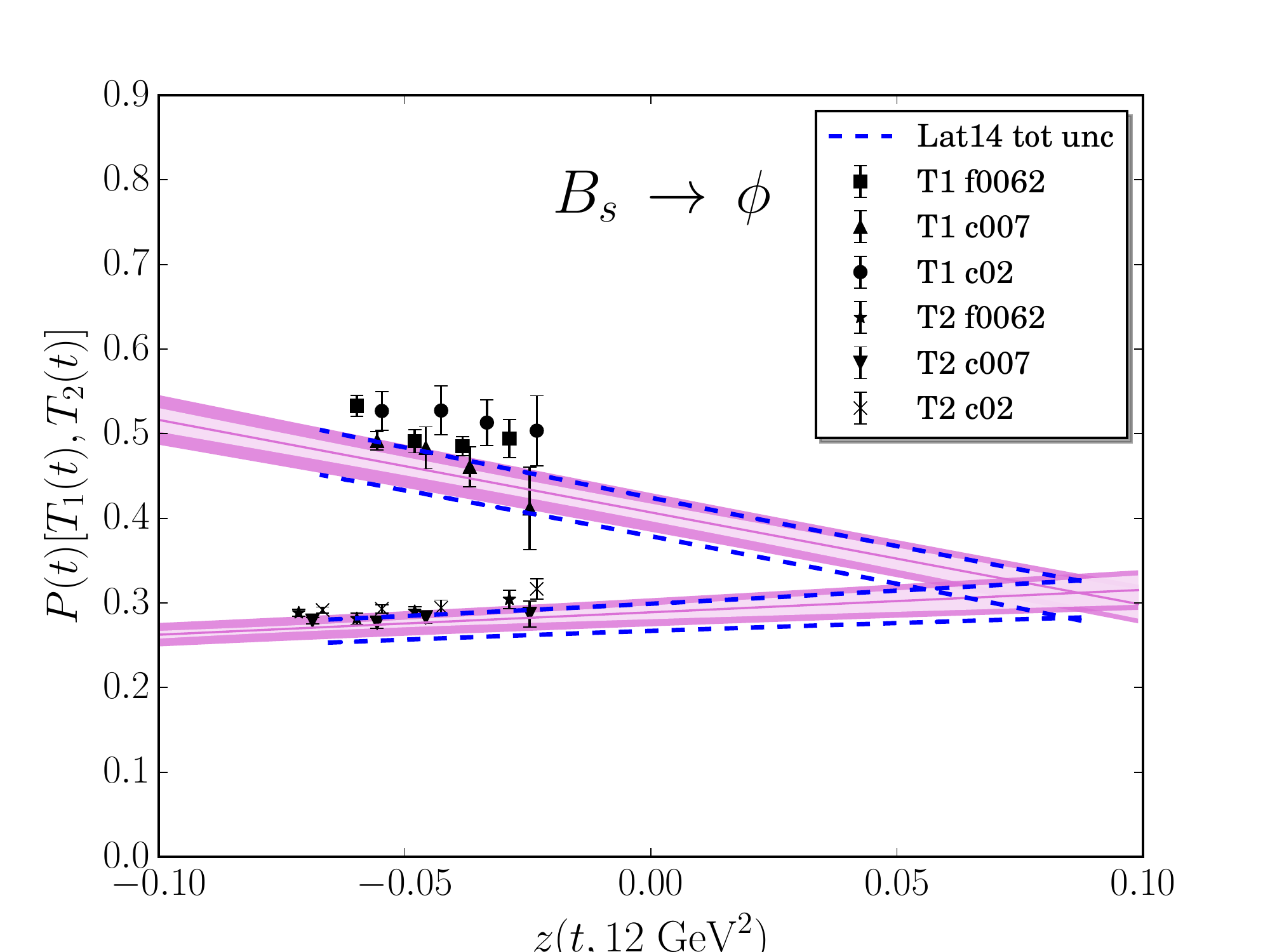}
\includegraphics[width=0.495\textwidth]{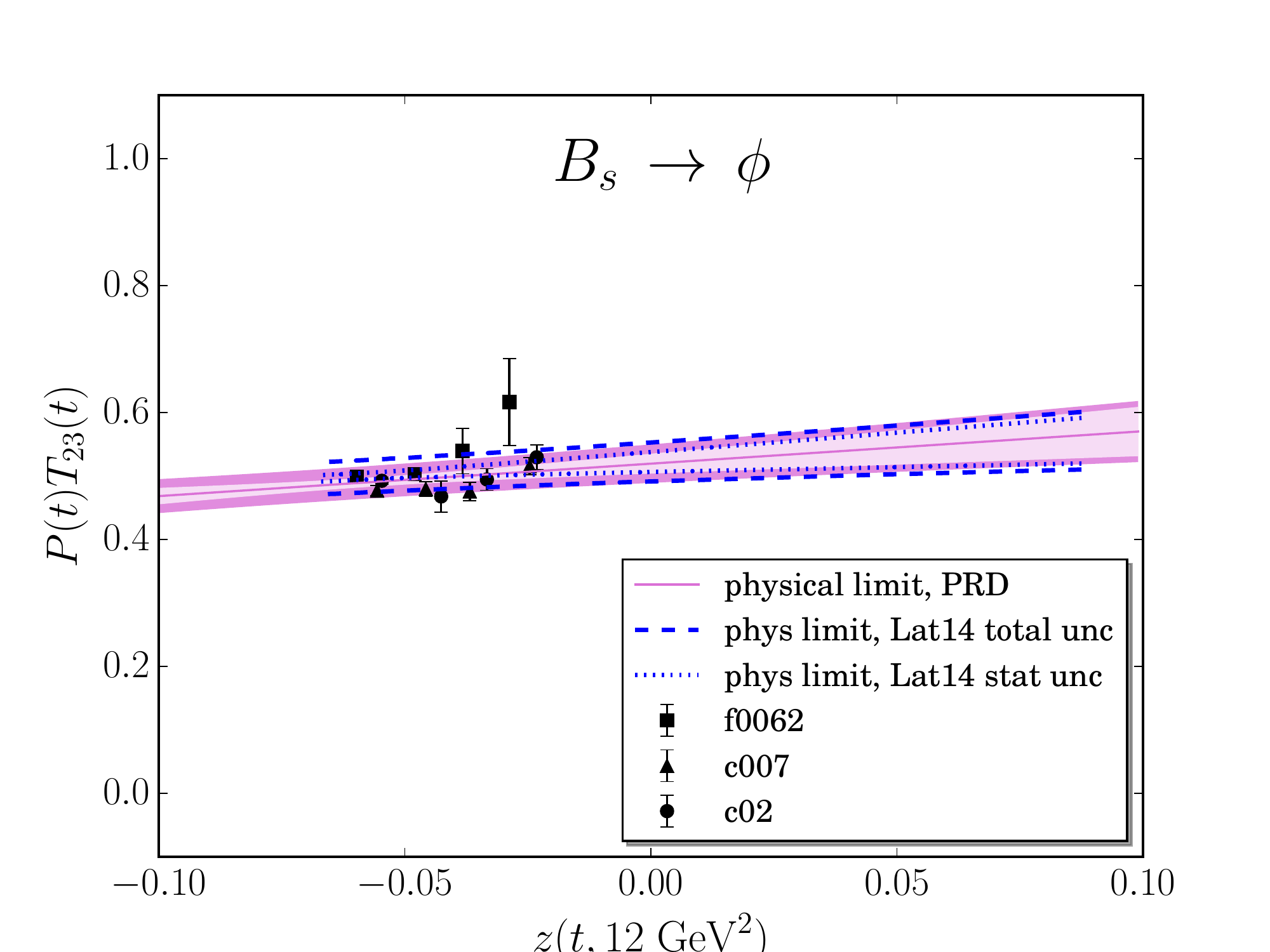}
\end{center}
\caption{\label{fig:pff_ss_hybrid}Comparison of $B_s\to \phi$ form
  factor fits presented here (blue dashed and dotted lines) to
  published fits \cite{Horgan:2013hoa} (pink bands). Black symbols are
  LQCD results at unphysical quark masses, whereas curves and bands
  are extrapolated to the physical limit.}
\end{figure}

\begin{figure}
\begin{center}
\includegraphics[width=0.495\textwidth]{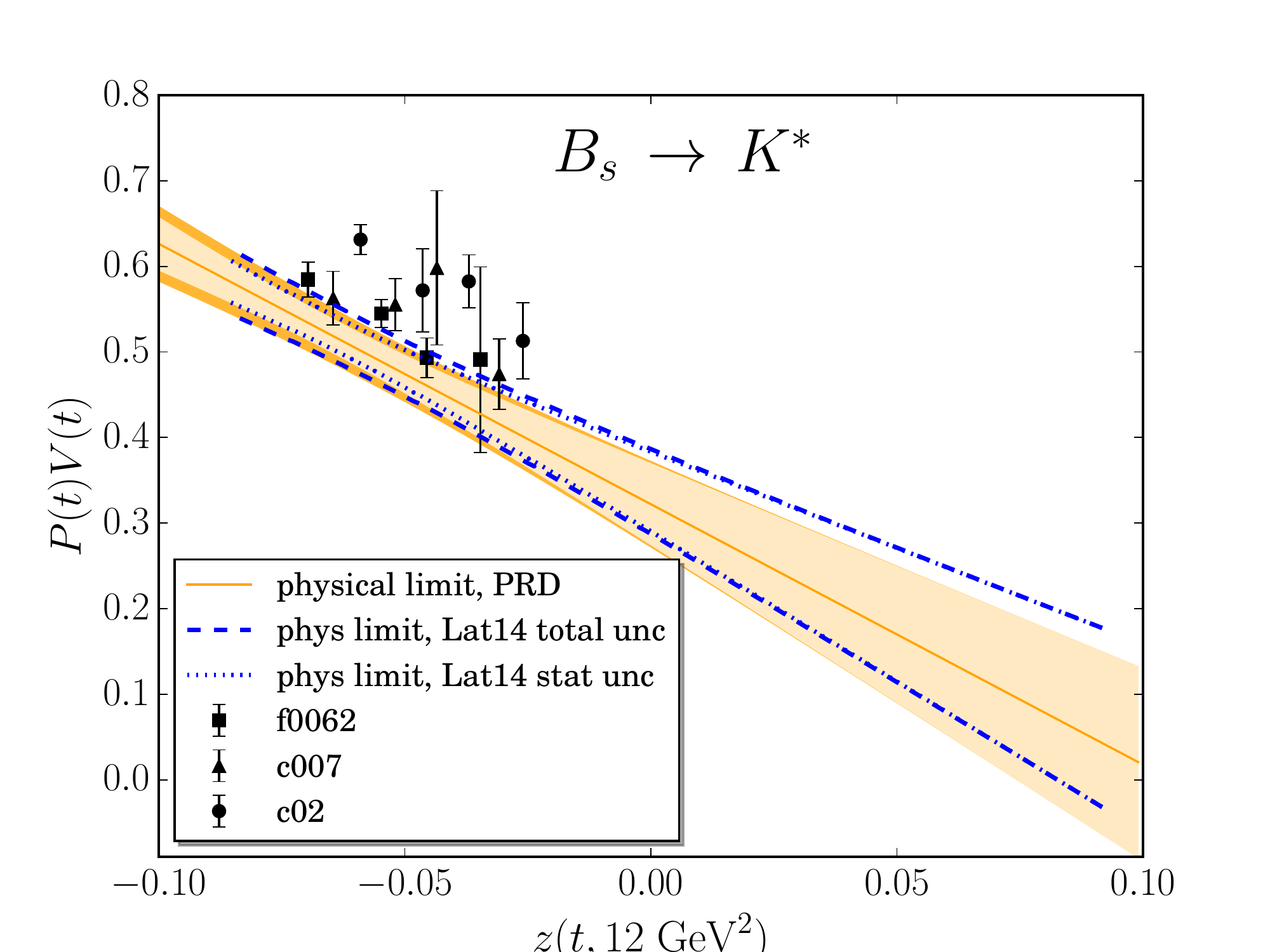}
\includegraphics[width=0.495\textwidth]{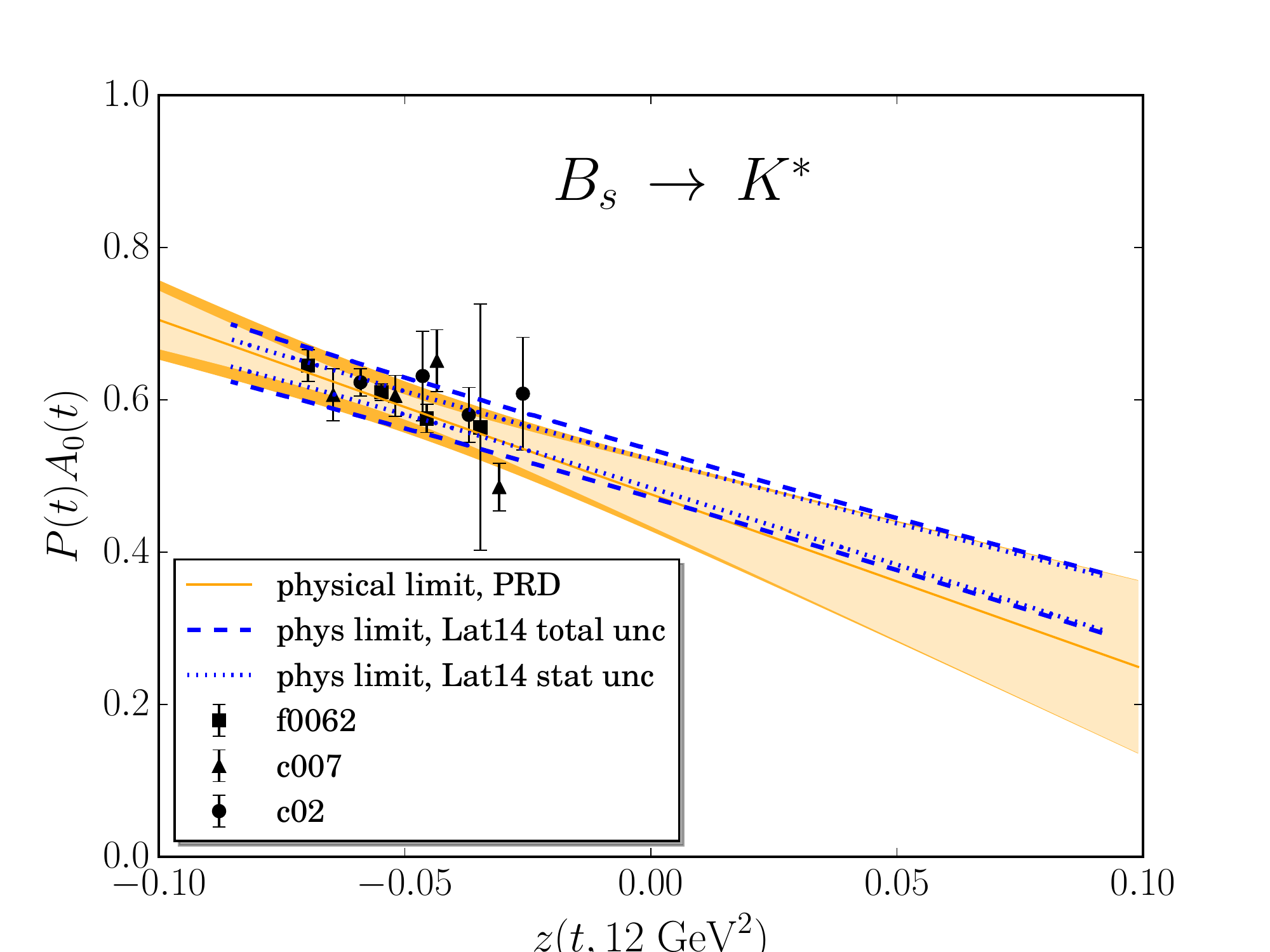}
\includegraphics[width=0.495\textwidth]{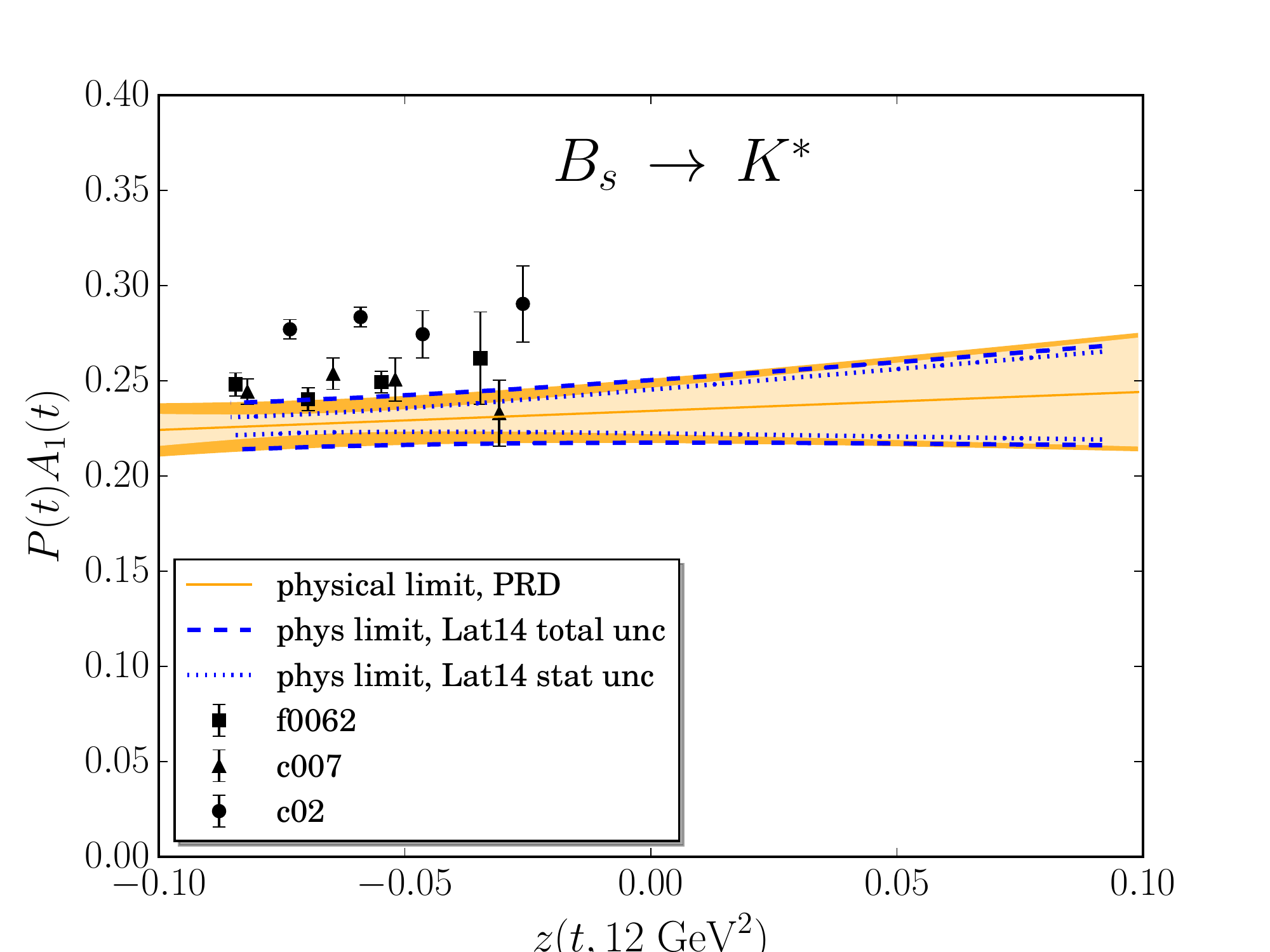}
\includegraphics[width=0.495\textwidth]{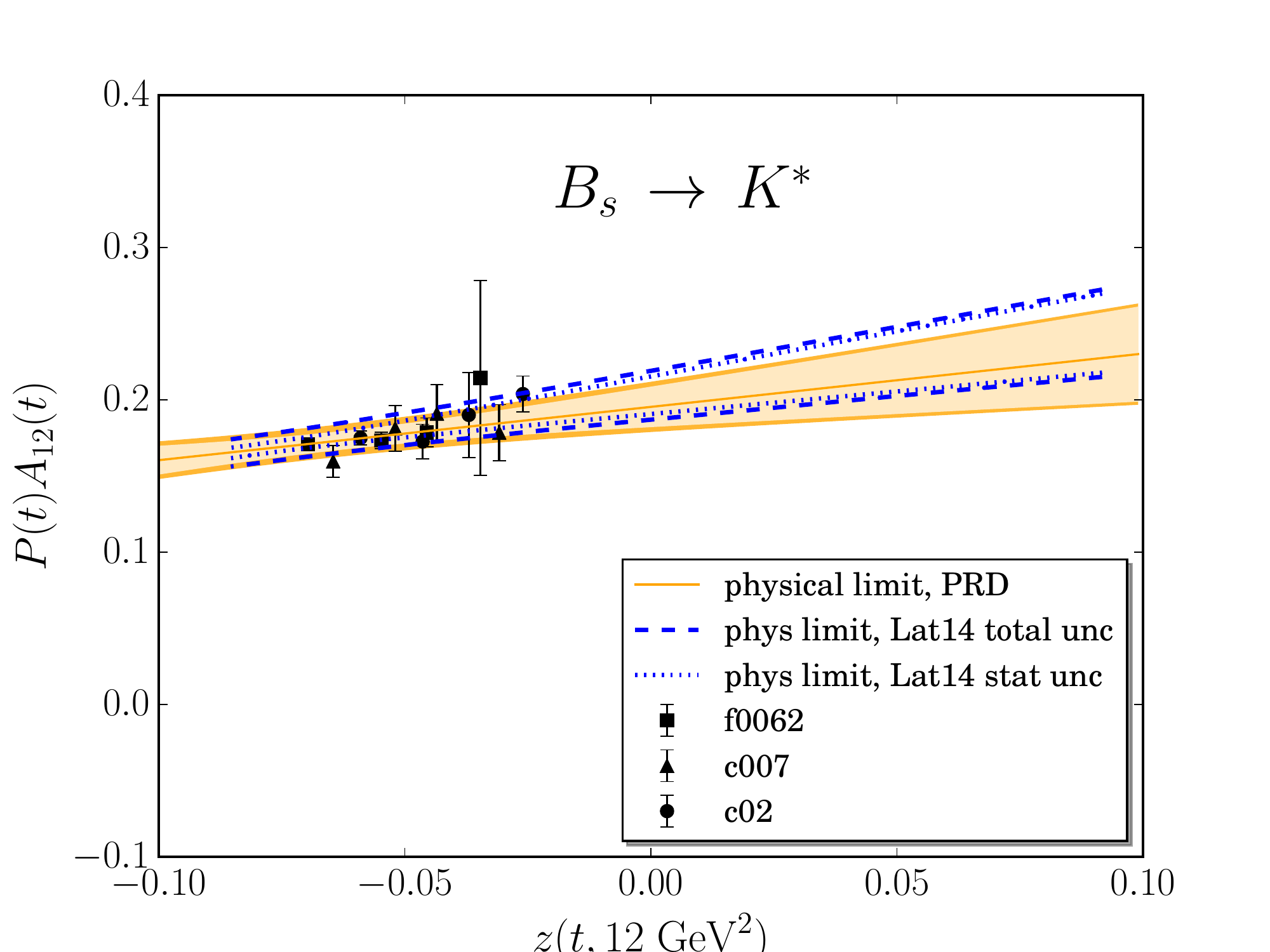}
\includegraphics[width=0.495\textwidth]{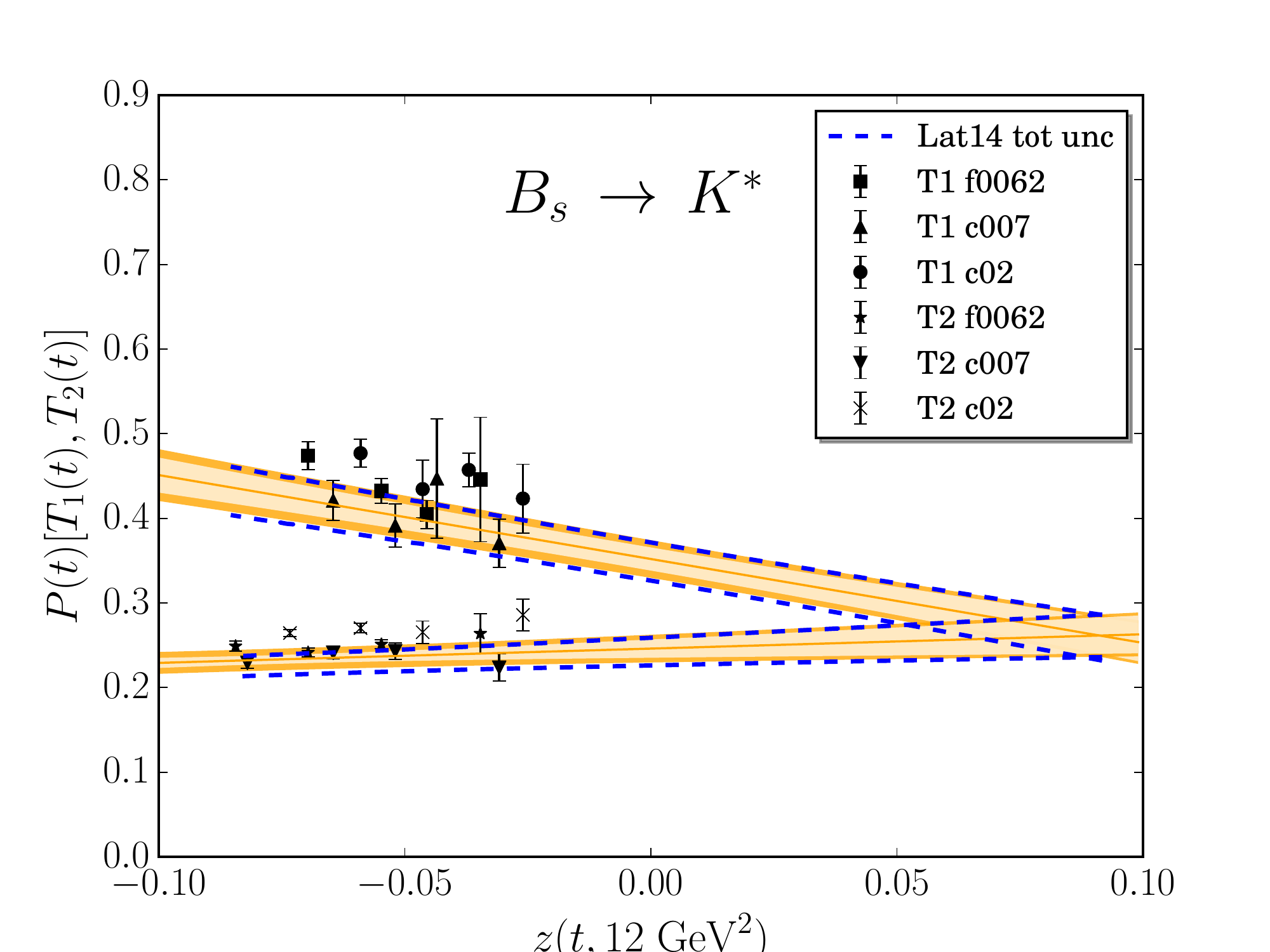}
\includegraphics[width=0.495\textwidth]{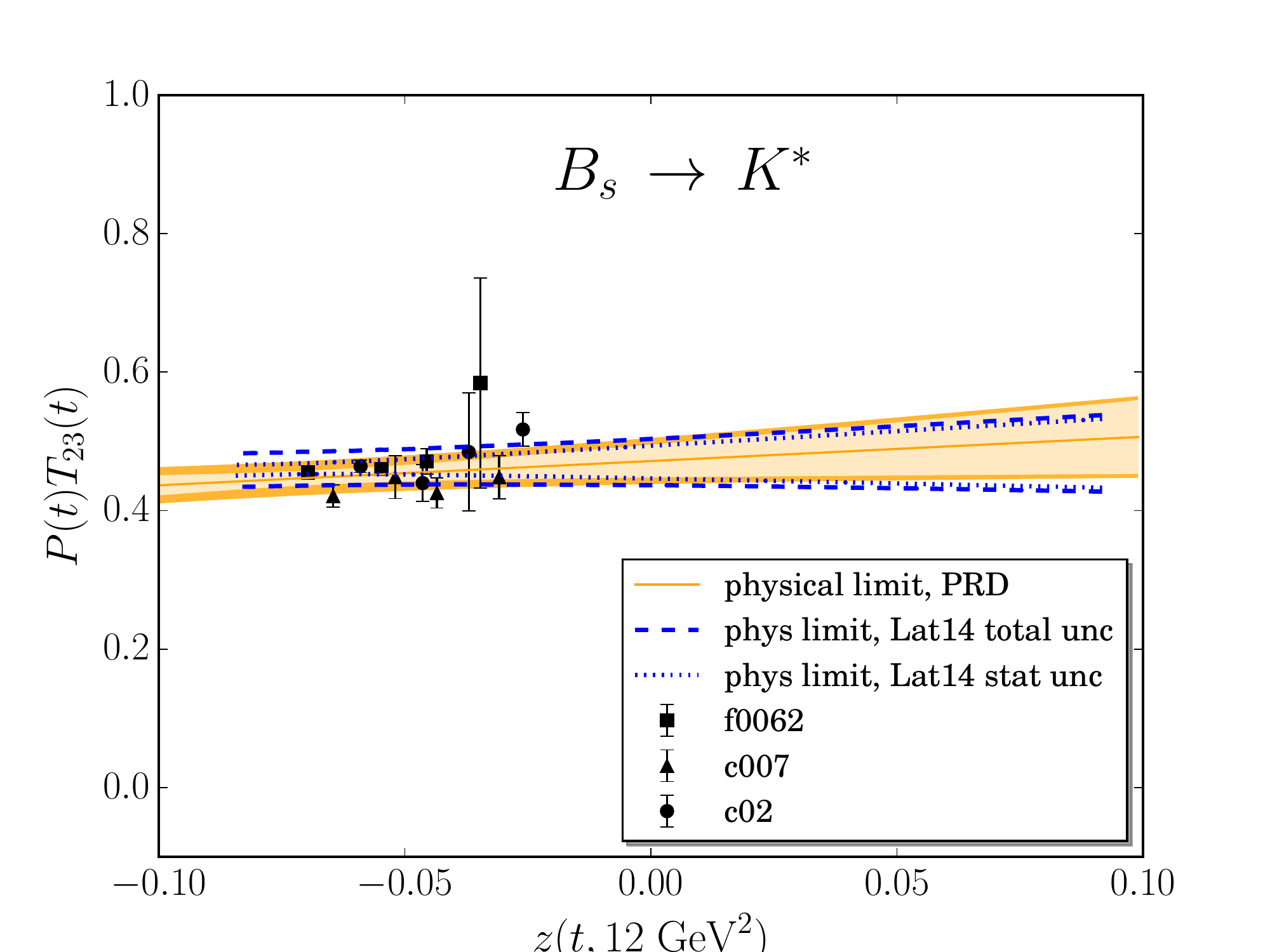}
\end{center}
\caption{\label{fig:pff_ls_hybrid}Comparison of $B_s\to K^*$ form factor
  fits presented here (blue dashed and dotted lines) compared to
  published fits \cite{Horgan:2013hoa} (orange bands).  Black symbols are
  LQCD results at unphysical quark masses, whereas curves and bands
  are extrapolated to the physical limit.}
\end{figure}

Figures \ref{fig:pff_sl_hybrid}, \ref{fig:pff_ss_hybrid}, and
\ref{fig:pff_ls_hybrid} show the fit results presented here compared to our
published results \cite{Horgan:2013hoa}.  The most visible difference in the
Figures is the reduction of uncertainty in $A_0$ at low values of $q^2$
(positive $z$) due to the constraint that it equals $A_{12}$ at $q^2=0$ up to
a known multiplicative factor.  The inclusion of the $q^2= t_-$ constraints
ensures endpoint relations are accurately satisfied \cite{Hiller:2013cza}.
Together these improvements to the fits yield slightly more precise and
accurate predictions for several observables.  Table~\ref{tab:ffpoints} gives
values for each of the form factors at a few reference kinematic points.

\begin{table}
\centering
\begin{tabular}{c|d{10}|rrrrrrr}\hline\hline
$p$ & \multicolumn{1}{c}{value}  & $C_{p,a_0^V}$  & $C_{p,a_1^V}$  & $C_{p,a_0^{A_0}}$  & $C_{p,a_1^{A_0}}$  & $C_{p,a_0^{A_1}}$  & $C_{p,a_1^{A_1}}$  & $C_{p,a_0^{A_{12}}}$  \\ \hline
$a_0^V$ & 0.4975(667) \\
$a_1^V$ & -2.015(916) & $0.8590$ \\
$a_0^{A_0}$ & 0.5023(370) & $0.0000$ & $0.0000$ \\
$a_1^{A_0}$ & -1.608(447) & $0.0000$ & $0.0000$ & $0.6673$ \\
$a_0^{A_1}$ & 0.2848(233) & $0.0000$ & $0.0000$ & $0.0002$ & $-0.0401$ \\
$a_1^{A_1}$ & 0.191(280) & $0.0000$ & $0.0000$ & $0.0000$ & $0.0104$ & $0.9483$ \\
$a_0^{A_{12}}$ & 0.2195(238) & $0.0000$ & $0.0000$ & $0.9043$ & $0.9091$ & $0.0073$ & $-0.0019$ \\
$a_1^{A_{12}}$ & 0.332(300) & $0.0000$ & $0.0000$ & $0.8878$ & $0.9272$ & $-0.0487$ & $0.0126$ & $0.9756$ \\
 \hline \hline
\end{tabular}
\caption{\label{tab:av_sl}Fit parameters and correlation matrix
  elements for $B\to K^*$ form factors $V$, $A_0$, $A_1$, and $A_{12}$.}
\end{table}

\begin{table}
\centering
\begin{tabular}{c|d{10}|rrrrr}\hline\hline
$p$ & \multicolumn{1}{c}{value}  & $C_{p,a_0^{T_1}}$  & $C_{p,a_1^{T_1}}$  & $C_{p,a_0^{T_2}}$  & $C_{p,a_1^{T_2}}$  & $C_{p,a_0^{T_{23}}}$  \\ \hline
$a_0^{T_1}$ & 0.4197(241) \\
$a_1^{T_1}$ & -1.363(259) & $0.5005$ \\
$a_0^{T_2}$ & 0.27997(1948) & $0.8503$ & $0.8015$ \\
$a_1^{T_2}$ & 0.117(236) & $0.8205$ & $0.8364$ & $0.9324$ \\
$a_0^{T_{23}}$ & 0.5235(451) & $0.0064$ & $-0.0022$ & $0.0117$ & $-0.0061$ \\
$a_1^{T_{23}}$ & -0.271(579) & $-0.0341$ & $0.0118$ & $-0.0628$ & $0.0325$ & $0.9520$
\\ \hline \hline
\end{tabular}
\caption{\label{tab:t_sl}Fit parameters and correlation matrix
  elements for $B\to K^*$ form factors $T_1$, $T_2$, and $T_{23}$.}
\end{table}

\begin{table}
\centering
\begin{tabular}{c|d{10}|rrrrrrr}\hline\hline
$p$ & \multicolumn{1}{c}{value}  & $C_{p,a_0^V}$  & $C_{p,a_1^V}$  & $C_{p,a_0^{A_0}}$  & $C_{p,a_1^{A_0}}$  & $C_{p,a_0^{A_1}}$  & $C_{p,a_1^{A_1}}$  & $C_{p,a_0^{A_{12}}}$  \\ \hline
$a_0^V$ & 0.4525(303) \\
$a_1^V$ & -2.399(496) & $0.8044$ \\
$a_0^{A_0}$ & 0.52935(1633) & $0.0000$ & $0.0000$ \\
$a_1^{A_0}$ & -1.592(243) & $0.0000$ & $0.0000$ & $0.7791$ \\
$a_0^{A_1}$ & 0.28283(1107) & $0.0000$ & $0.0000$ & $-0.0097$ & $-0.0774$ \\
$a_1^{A_1}$ & 0.1238(1501) & $0.0000$ & $0.0000$ & $-0.0006$ & $-0.0047$ & $0.9353$ \\
$a_0^{A_{12}}$ & 0.20661(978) & $0.0000$ & $0.0000$ & $0.9322$ & $0.9364$ & $0.0022$ & $0.0001$ \\
$a_1^{A_{12}}$ & 0.4763(1473) & $0.0000$ & $0.0000$ & $0.9111$ & $0.9598$ & $-0.0904$ & $-0.0055$ & $0.9690$ \\
\hline \hline
\end{tabular}
\caption{\label{tab:av_ss}Fit parameters and correlation matrix
  elements for $B_s\to \phi$ form factors $V$, $A_0$, $A_1$, and $A_{12}$.}
\end{table}

\begin{table}
\centering
\begin{tabular}{c|d{10}|rrrrr}\hline\hline
$p$ & \multicolumn{1}{c}{value}  & $C_{p,a_0^{T_1}}$  & $C_{p,a_1^{T_1}}$  & $C_{p,a_0^{T_2}}$  & $C_{p,a_1^{T_2}}$  & $C_{p,a_0^{T_{23}}}$  \\ \hline
$a_0^{T_1}$ & 0.40160(1057) \\
$a_1^{T_1}$ & -1.1340(1325) & $0.3041$ \\
$a_0^{T_2}$ & 0.28297(752) & $0.7021$ & $0.6263$ \\
$a_1^{T_2}$ & 0.2487(975) & $0.6767$ & $0.6467$ & $0.9273$ \\
$a_0^{T_{23}}$ & 0.52193(1566) & $0.0202$ & $0.0104$ & $0.0351$ & $0.0080$ \\
$a_1^{T_{23}}$ & 0.384(236) & $-0.0689$ & $-0.0357$ & $-0.1199$ & $-0.0273$ & $0.9455$
\\ \hline \hline
\end{tabular}
\caption{\label{tab:t_ss}Fit parameters and correlation matrix
  elements for $B_s\to \phi$ form factors $T_1$, $T_2$, and $T_{23}$.}
\end{table}

\begin{table}
\centering
\begin{tabular}{c|d{10}|rrrrrrr}\hline\hline
$p$ & \multicolumn{1}{c}{value}  & $C_{p,a_0^V}$  & $C_{p,a_1^V}$  & $C_{p,a_0^{A_0}}$  & $C_{p,a_1^{A_0}}$  & $C_{p,a_0^{A_1}}$  & $C_{p,a_1^{A_1}}$  & $C_{p,a_0^{A_{12}}}$  \\ \hline
$a_0^V$ & 0.3367(466) \\
$a_1^V$ & -2.879(661) & $0.9037$ \\
$a_0^{A_0}$ & 0.50344(1856) & $0.0000$ & $0.0000$ \\
$a_1^{A_0}$ & -1.855(235) & $0.0000$ & $0.0000$ & $0.5757$ \\
$a_0^{A_1}$ & 0.23390(1151) & $0.0000$ & $0.0000$ & $-0.0173$ & $-0.0681$ \\
$a_1^{A_1}$ & 0.0907(1316) & $0.0000$ & $0.0000$ & $0.0140$ & $0.0554$ & $0.9119$ \\
$a_0^{A_{12}}$ & 0.20303(1233) & $0.0000$ & $0.0000$ & $0.8777$ & $0.8804$ & $-0.0175$ & $0.0142$ \\
$a_1^{A_{12}}$ & 0.4442(1509) & $0.0000$ & $0.0000$ & $0.8480$ & $0.9095$ & $-0.0785$ & $0.0638$ & $0.9637$ 
 \\ \hline \hline
\end{tabular}
\caption{\label{tab:av_ls}Fit parameters and correlation matrix
  elements for $B_s\to K^*$ form factors $V$, $A_0$, $A_1$, and $A_{12}$. }
\end{table}

\begin{table}
\centering
\begin{tabular}{c|d{10}|rrrrr}\hline\hline
$p$ & \multicolumn{1}{c}{value}  & $C_{p,a_0^{T_1}}$  & $C_{p,a_1^{T_1}}$  & $C_{p,a_0^{T_2}}$  & $C_{p,a_1^{T_2}}$  & $C_{p,a_0^{T_{23}}}$  \\ \hline
$a_0^{T_1}$ & 0.34878(1423) \\
$a_1^{T_1}$ & -0.9807(1783) & $0.1851$ \\
$a_0^{T_2}$ & 0.24243(1063) & $0.6786$ & $0.6929$ \\
$a_1^{T_2}$ & 0.2051(1220) & $0.6440$ & $0.7236$ & $0.9296$ \\
$a_0^{T_{23}}$ & 0.4701(237) & $-0.0026$ & $0.0018$ & $-0.0052$ & $0.0044$ \\
$a_1^{T_{23}}$ & 0.138(291) & $-0.0356$ & $0.0244$ & $-0.0703$ & $0.0595$ &
$0.9468$\\
\hline\hline
\end{tabular}
\caption{\label{tab:t_ls}Fit parameters and correlation matrix
  elements for $B_s\to K^*$ form factors $T_1$, $T_2$, and $T_{23}$.}
\end{table}

\begin{table}
\centering
\begin{tabular}{cd{6}d{6}rrd{6}rd{6}} \hline\hline
\multicolumn{8}{c}{$B\to K^*$} \\
$q^2$ & \multicolumn{1}{c}{$V$} & \multicolumn{1}{c}{$A_0$} & 
\multicolumn{1}{c}{$A_1$} & \multicolumn{1}{c}{$A_{12}$} & 
\multicolumn{1}{c}{$T_1$} & \multicolumn{1}{c}{$T_2$} & 
\multicolumn{1}{c}{$T_{23}$} \\ \hline 
$q_{\mathrm{max}}^2$ & 1.92(15) & 1.90(13) & 0.620(35) & 0.444(25) & 1.54(10) & 0.623(35) & 1.255(71) \\
16 & 1.28(11) & 1.280(90) & 0.523(36) & 0.389(30) & 1.049(70) & 0.520(33) & 1.010(67) \\
12 & 0.84(12) & 0.861(77) & 0.440(42) & 0.339(40) & 0.711(54) & 0.433(37) & 0.809(81) \\
0 & 0.31(15) & 0.351(74) & 0.303(51) & 0.251(53) & 0.291(44) & 0.291(44) & 0.50(10) \\ \hline
\multicolumn{8}{c}{$B_s\to \phi$} \\
$q^2$ & \multicolumn{1}{c}{$V$} & \multicolumn{1}{c}{$A_0$} & 
\multicolumn{1}{c}{$A_1$} & \multicolumn{1}{c}{$A_{12}$} & 
\multicolumn{1}{c}{$T_1$} & \multicolumn{1}{c}{$T_2$} & 
\multicolumn{1}{c}{$T_{23}$} \\ \hline 
$q_{\mathrm{max}}^2$ & 1.74(10) & 1.856(98) & 0.624(32) & 0.396(21) & 1.351(75) & 0.605(31) & 1.128(58) \\
16 & 1.195(73) & 1.325(71) & 0.529(29) & 0.359(20) & 0.979(54) & 0.521(27) & 0.966(51) \\
12 & 0.767(64) & 0.907(53) & 0.439(28) & 0.321(22) & 0.680(38) & 0.439(25) & 0.810(47) \\
0 & 0.244(71) & 0.391(40) & 0.294(28) & 0.248(26) & 0.303(23) & 0.305(22) & 0.555(45) \\ \hline
\multicolumn{8}{c}{$B_s\to K^*$} \\
$q^2$ & \multicolumn{1}{c}{$V$} & \multicolumn{1}{c}{$A_0$} & 
\multicolumn{1}{c}{$A_1$} & \multicolumn{1}{c}{$A_{12}$} & 
\multicolumn{1}{c}{$T_1$} & \multicolumn{1}{c}{$T_2$} & 
\multicolumn{1}{c}{$T_{23}$} \\ \hline 
$q_{\mathrm{max}}^2$ & 1.99(13) & 2.35(13) & 0.582(32) & 0.424(23) & 1.470(98) & 0.578(31) & 1.180(63) \\
16 & 1.035(78) & 1.355(77) & 0.451(26) & 0.363(22) & 0.890(56) & 0.459(26) & 0.910(52) \\
12 & 0.584(86) & 0.884(55) & 0.370(26) & 0.321(25) & 0.605(39) & 0.383(25) & 0.743(53) \\
0 & 0.07(10) & 0.335(39) & 0.242(26) & 0.243(29) & 0.259(27) & 0.261(25) & 0.483(55) \\ \hline \hline
\end{tabular}
\caption{\label{tab:ffpoints}Form factor values (and total estimated 
uncertainties) at several reference values
of $q^2$ (in GeV${}^2$) using updated fits to our lattice results.}
\end{table}

\end{document}